%% file: main.tex
\pdfoutput=1
\documentclass{article}

\usepackage[letterpaper,margin=1in]{geometry}
\usepackage{times}

\usepackage{hyperref}
\usepackage{url}
\usepackage{booktabs}
\usepackage{graphicx}
\usepackage{float}
\usepackage{amsmath,amssymb}
\usepackage{xcolor}
\usepackage{natbib}

\input{macros_style}

\input{macros_numbers}
\title{What and Where in Identifying and Quantifying\\
       Conventions in Humans and AI Agents}

\author{Makoto Fukushima\\Honda Research Institute Japan \and Tim Puphal\\Honda Research Institute Europe}

\newcommand{\suppfig}{Appendix Fig.}
\newcommand{\supptab}{Appendix Table}
\renewcommand{\maintext}{}
\renewcommand{\Maintext}{}
\renewcommand{\releasesentence}{The analysis pipelines, the audit JSONs, the generated tables and figures, and the derived per-play records for the agent conditions will be released with the published version.}
\renewcommand{\releaseanon}{}
\newcommand{\Suppmat}{The appendix}

\newcommand{\appendixCarriesMovedFigs}{}

\begin{document}
\maketitle

\begin{abstract}
\input{sections/00_abstract}
\end{abstract}

\input{sections_aamas/01_intro}
\input{sections/01b_related}
\input{sections_aamas/02_setup}
\input{sections_aamas/03_what}
\input{sections_aamas/04_when}
\input{sections_aamas/06_worth}
\input{sections_aamas/07_discussion}

\bibliographystyle{plainnat}
\bibliography{refs}

\appendix

\input{appendix/A_metric}

\input{appendix/B_mixed}

\input{appendix/C_seeds}

\input{appendix/C_interventions}
\input{appendix/D_corpus}
\input{appendix/E_depth}
\input{appendix/F_repro}
\input{appendix/G_clones}

\end{document}

%% file: macros_style.tex
\newcommand{\pp}{\,\text{pp}}

\newcommand{\paperone}{\citep{paper1}}

\providecommand{\metriccite}{\citep{paper1}}

\DeclareUnicodeCharacter{2014}{---}
\DeclareUnicodeCharacter{2192}{\ensuremath{\rightarrow}}

\def\DotRangeAux#1..#2\relax{#1 to #2}
\newcommand{\dotrange}[1]{\expandafter\DotRangeAux#1\relax}
\def\HyRangeAux#1-#2\relax{#1--#2}
\newcommand{\hyrange}[1]{\expandafter\HyRangeAux#1\relax}
\def\SlashOfAux#1/#2\relax{#1 of #2}
\newcommand{\slashof}[1]{\expandafter\SlashOfAux#1\relax}
\def\BeforeParenAux#1 (#2\relax{#1}
\newcommand{\beforeparen}[1]{\expandafter\BeforeParenAux#1\relax}
\def\PtOfAux#1~[#2\relax{#1}
\newcommand{\ptof}[1]{\expandafter\PtOfAux#1~[\relax}
\providecommand{\maintext}{main-text\space}
\providecommand{\Maintext}{Main-text\space}
\providecommand{\releasesentence}{The camera-ready release comprises the analysis pipelines, the audit JSONs, the generated tables and figures, and the derived per-play records for the agent conditions.}
\providecommand{\releaseanon}{ Paths, repository names and archive identifiers are withheld from this submission for anonymity and appear in the camera-ready.}

\AtBeginDocument{%
  \renewcommand{\gapOBLOne}{+1.57~[+1.33, +1.81]}%
  \renewcommand{\gapOBLTwo}{+12.40~[+12.10, +12.71]}%
  \renewcommand{\gapOBLFive}{+21.68~[+21.34, +22.02]}%
  \renewcommand{\gapOBLOneCross}{+1.22~[+0.98, +1.45]}%
}
\newcommand{\poneTierOneStrictGames}{424}
\newcommand{\poneTierOneStrictPlays}{9{,}011}
\newcommand{\poneTierOneStrictGap}{+26.18}

%% file: macros_numbers.tex
\newcommand{\gapOBLOne}{+1.57~[+1.33, +1.80]}%
\newcommand{\gapOBLOnePt}{+1.57}%
\newcommand{\scoreOBLOne}{21.09}%
\newcommand{\gapOBLTwo}{+12.40~[+12.11, +12.69]}%
\newcommand{\gapOBLTwoPt}{+12.40}%
\newcommand{\gapOBLFive}{+21.68~[+21.34, +22.03]}%
\newcommand{\gapOBLFivePt}{+21.68}%
\newcommand{\scoreOBLFive}{24.16}%
\newcommand{\gapOBLOneCross}{+1.22~[+0.98, +1.45]}%
\newcommand{\tFourscoreOnexTwo}{21.37}%
\newcommand{\tFourgapLoOnexTwo}{+1.0~[+0.7, +1.3]}%
\newcommand{\tFourgapHiOnexTwo}{+8.4~[+8.0, +8.9]}%
\newcommand{\tFourgapHiOnexFour}{+9.0~[+8.6, +9.4]}%
\newcommand{\tFourscoreOnexFive}{19.33}%
\newcommand{\tFourgapLoOnexFive}{+0.1~[-0.2, +0.5]}%
\newcommand{\tFourgapLoTwoxFive}{+13.4~[+13.0, +13.9]}%
\newcommand{\tFourgapHiTwoxFive}{+14.0~[+13.6, +14.5]}%
\newcommand{\tSevenOBLTwoxOBLThreeScore}{23.32}%
\newcommand{\tSevenOBLTwoxOBLThreeScoreSd}{0.20}%
\newcommand{\tSevenOBLTwoxOBLThreeGapLow}{+13.1}%
\newcommand{\tSevenOBLTwoxOBLThreeGapLowSd}{0.7}%
\newcommand{\tSevenOBLTwoxOBLThreeGapHigh}{+13.4}%
\newcommand{\tSevenOBLTwoxOBLThreeGapHighSd}{0.8}%
\newcommand{\tSevenOBLTwoxOBLThreeDeltaPt}{+0.28}%
\newcommand{\tSevenOBLTwoxOBLThreeDeltaCiLo}{-0.12}%
\newcommand{\tSevenOBLTwoxOBLThreeDeltaCiHi}{+0.67}%
\newcommand{\tSevenOBLTwoxOBLThreeSeedShare}{42}%
\newcommand{\tSevenOBLTwoxOBLFourScore}{23.15}%
\newcommand{\tSevenOBLTwoxOBLFourScoreSd}{0.31}%
\newcommand{\tSevenOBLTwoxOBLFourGapLow}{+13.2}%
\newcommand{\tSevenOBLTwoxOBLFourGapLowSd}{1.1}%
\newcommand{\tSevenOBLTwoxOBLFourGapHigh}{+13.9}%
\newcommand{\tSevenOBLTwoxOBLFourGapHighSd}{0.8}%
\newcommand{\tSevenOBLTwoxOBLFourDelta}{+0.70~[+0.26, +1.14]}%
\newcommand{\tSevenOBLTwoxOBLFourDeltaPt}{+0.70}%
\newcommand{\tSevenOBLTwoxOBLFourDeltaCiLo}{+0.26}%
\newcommand{\tSevenOBLTwoxOBLFourDeltaCiHi}{+1.14}%
\newcommand{\tSevenOBLTwoxOBLFourSeedShare}{52}%
\newcommand{\tSevenOBLTwoxOBLFiveScore}{22.84}%
\newcommand{\tSevenOBLTwoxOBLFiveScoreSd}{0.30}%
\newcommand{\tSevenOBLTwoxOBLFiveGapLow}{+12.9}%
\newcommand{\tSevenOBLTwoxOBLFiveGapLowSd}{1.1}%
\newcommand{\tSevenOBLTwoxOBLFiveGapHigh}{+13.6}%
\newcommand{\tSevenOBLTwoxOBLFiveGapHighSd}{0.8}%
\newcommand{\tSevenOBLTwoxOBLFiveDelta}{+0.64~[+0.22, +1.07]}%
\newcommand{\tSevenOBLTwoxOBLFiveDeltaPt}{+0.64}%
\newcommand{\tSevenOBLTwoxOBLFiveDeltaCiLo}{+0.22}%
\newcommand{\tSevenOBLTwoxOBLFiveDeltaCiHi}{+1.07}%
\newcommand{\tSevenOBLTwoxOBLFiveSeedShare}{48}%
\newcommand{\tSevenOBLThreexOBLFourScore}{23.74}%
\newcommand{\tSevenOBLThreexOBLFourScoreSd}{0.26}%
\newcommand{\tSevenOBLThreexOBLFourGapLow}{+17.0}%
\newcommand{\tSevenOBLThreexOBLFourGapLowSd}{1.2}%
\newcommand{\tSevenOBLThreexOBLFourGapHigh}{+16.9}%
\newcommand{\tSevenOBLThreexOBLFourGapHighSd}{1.0}%
\newcommand{\tSevenOBLThreexOBLFourDeltaPt}{-0.05}%
\newcommand{\tSevenOBLThreexOBLFourDeltaCiLo}{-0.37}%
\newcommand{\tSevenOBLThreexOBLFourDeltaCiHi}{+0.27}%
\newcommand{\tSevenOBLThreexOBLFourSeedShare}{10}%
\newcommand{\tSevenOBLThreexOBLFiveScore}{23.67}%
\newcommand{\tSevenOBLThreexOBLFiveScoreSd}{0.35}%
\newcommand{\tSevenOBLThreexOBLFiveGapLow}{+17.3}%
\newcommand{\tSevenOBLThreexOBLFiveGapLowSd}{1.2}%
\newcommand{\tSevenOBLThreexOBLFiveGapHigh}{+17.0}%
\newcommand{\tSevenOBLThreexOBLFiveGapHighSd}{1.3}%
\newcommand{\tSevenOBLThreexOBLFiveDeltaPt}{-0.25}%
\newcommand{\tSevenOBLThreexOBLFiveDeltaCiLo}{-0.67}%
\newcommand{\tSevenOBLThreexOBLFiveDeltaCiHi}{+0.16}%
\newcommand{\tSevenOBLThreexOBLFiveSeedShare}{45}%
\newcommand{\tSevenOBLFourxOBLFiveScore}{23.80}%
\newcommand{\tSevenOBLFourxOBLFiveScoreSd}{0.30}%
\newcommand{\tSevenOBLFourxOBLFiveGapLow}{+19.2}%
\newcommand{\tSevenOBLFourxOBLFiveGapLowSd}{1.5}%
\newcommand{\tSevenOBLFourxOBLFiveGapHigh}{+18.7}%
\newcommand{\tSevenOBLFourxOBLFiveGapHighSd}{1.6}%
\newcommand{\tSevenOBLFourxOBLFiveDelta}{-0.45~[-0.90, -0.00]}%
\newcommand{\tSevenOBLFourxOBLFiveDeltaPt}{-0.45}%
\newcommand{\tSevenOBLFourxOBLFiveDeltaCiLo}{-0.90}%
\newcommand{\tSevenOBLFourxOBLFiveDeltaCiHi}{-0.00}%
\newcommand{\tSevenOBLFourxOBLFiveSeedShare}{48}%
\newcommand{\tSevenOBLOnexOBLFiveScore}{18.62}%
\newcommand{\tSevenOBLOnexOBLFiveScoreSd}{0.79}%
\newcommand{\tSevenOBLOnexOBLFiveGapLow}{+0.2}%
\newcommand{\tSevenOBLOnexOBLFiveGapLowSd}{0.5}%
\newcommand{\tSevenOBLOnexOBLFiveGapHigh}{+8.0}%
\newcommand{\tSevenOBLOnexOBLFiveGapHighSd}{0.9}%
\newcommand{\tSevenOBLOnexOBLFiveDelta}{+7.77~[+7.33, +8.20]}%
\newcommand{\tSevenOBLOnexOBLFiveDeltaPt}{+7.77}%
\newcommand{\tSevenOBLOnexOBLFiveDeltaCiLo}{+7.33}%
\newcommand{\tSevenOBLOnexOBLFiveDeltaCiHi}{+8.20}%
\newcommand{\tSevenOBLOnexOBLFiveSeedShare}{59}%
\newcommand{\tSevenControlDeltaMin}{+6.3}%
\newcommand{\tSevenControlDeltaMax}{+10.3}%
\newcommand{\tSevenControlAllPositive}{25}%
\newcommand{\tSevenBlendTwotoFiveMaxAbsDelta}{0.70}%
\newcommand{\tSevenSelfplaySdOBLTwo}{0.61}%
\newcommand{\tSevenSelfplayMeanOBLTwo}{+12.34}%
\newcommand{\tSevenSelfplaySdOBLThree}{0.67}%
\newcommand{\tSevenSelfplaySdOBLFour}{1.27}%
\newcommand{\tSevenSelfplaySdOBLFive}{1.53}%
\newcommand{\tSevenSelfplayMeanOBLFive}{+21.11}%
\newcommand{\tSevenMonotoneSeeds}{5}%
\newcommand{\tSevenGamesTotal}{35000}%
\newcommand{\tEightOblFiveBaseScore}{24.10}%
\newcommand{\tEightOblFiveMaskBothScore}{22.07}%
\newcommand{\tEightOblFiveMaskBothDeltaScore}{-2.03~[-2.25, -1.81]}%
\newcommand{\tEightOblFiveMaskPZeroDeltaScore}{-0.92~[-1.10, -0.73]}%
\newcommand{\tEightOblFivePersistDeltaScore}{-5.77~[-6.04, -5.49]}%
\newcommand{\tEightOblFivePersistGap}{+3.5~[+3.17, +3.74]}%
\newcommand{\tEightOblTwoMaskBothDeltaScore}{-1.21~[-1.38, -1.04]}%
\newcommand{\tEightStaleShareBase}{34}%
\newcommand{\tEightStaleShareMask}{66}%
\newcommand{\tEightStaleGapBase}{+6.4~[+5.9, +6.9]}%
\newcommand{\tEightStaleGapMask}{+17.5~[+17.1, +18.0]}%
\newcommand{\tEightPersistDiscardShare}{40}%
\newcommand{\tEightPZeroMaskedPlayerGapBase}{+22.0~[+21.43, +22.48]}%
\newcommand{\tEightPZeroMaskedPlayerGapMask}{+15.5~[+14.94, +15.97]}%
\newcommand{\tEightPZeroUnmaskedPlayerGapBase}{+20.6~[+20.16, +21.13]}%
\newcommand{\tEightPZeroUnmaskedPlayerGapMask}{+20.1~[+19.56, +20.56]}%
\newcommand{\tNineaDiscOBLOne}{11,309}%
\newcommand{\tNineaEstOBLOne}{11,482}%
\newcommand{\tNineaTotalOBLOne}{22,791}%
\newcommand{\tNineaDiscOBLTwo}{60,573}%
\newcommand{\tNineaEstOBLTwo}{60,534}%
\newcommand{\tNineaTotalOBLTwo}{121,107}%
\newcommand{\tNineaDiscOBLThree}{61,228}%
\newcommand{\tNineaEstOBLThree}{61,487}%
\newcommand{\tNineaTotalOBLThree}{122,715}%
\newcommand{\tNineaDiscOBLFour}{61,761}%
\newcommand{\tNineaEstOBLFour}{61,842}%
\newcommand{\tNineaTotalOBLFour}{123,603}%
\newcommand{\tNineaDiscOBLFive}{62,038}%
\newcommand{\tNineaEstOBLFive}{61,996}%
\newcommand{\tNineaTotalOBLFive}{124,034}%
\newcommand{\tNineaDiscAll}{256,909}%
\newcommand{\tNineaEstAll}{257,341}%
\newcommand{\tNineaTotalAll}{514,250}%
\newcommand{\tNineRsqBackoffOBLOne}{0.058}%
\newcommand{\tNineRsqCartOBLOne}{0.069}%
\newcommand{\tNineCatsOBLFive}{105}%
\newcommand{\tNineRsqFiveOBLFive}{0.215}%
\newcommand{\tNineRsqBackoffOBLFive}{0.570}%
\newcommand{\tNineRsqCartOBLFive}{0.571}%
\newcommand{\tOneZeroGapOne}{+62.2~[+60.1, +64.0]}%
\newcommand{\tOneZeroGapOnePt}{+62.2}%
\newcommand{\tOneZeroGapTwo}{+62.2~[+60.0, +64.2]}%
\newcommand{\tOneZeroGapThree}{+56.8~[+56.4, +57.2]}%
\newcommand{\tOneZeroGapFour}{+55.6~[+55.1, +56.2]}%
\newcommand{\txTypesRange}{303--374}%
\newcommand{\txKNineZero}{OBL1=90, OBL2=74, OBL3=72, OBL4=70, OBL5=71}%
\newcommand{\txNminRange}{52--139}%
\newcommand{\txSdRange}{0.18--0.30}%
\newcommand{\txCovThreepp}{77--82\% (OBL2--5; OBL1 53\%)}%
\newcommand{\txStaleHintedOBLFive}{+6.4}%
\newcommand{\txTokenModulationOBLFive}{+54.4 at 0 tokens vs +55.5/+57.0 at 1--2/3+ (color-only newest, est half); Table-10 cells +54.6 / +55.6 / +56.8}%
\newcommand{\txNewestVsOlder}{12.7--32.5 pp (matched fresh-color C lag-1 cells, OBL2--5)}%
\newcommand{\txConcentration}{OBL1=5, OBL2=7, OBL3=7, OBL4=7, OBL5=8}%
\newcommand{\txHetChecked}{40}%
\newcommand{\txHetFlags}{18}%
\newcommand{\txHetHeadExcessRange}{1.5--2.9 pp (3--9\% of gap)}%
\newcommand{\txHetLargeExcessMaxShare}{0.87\%}%
\newcommand{\txHetOBLThreeDiscardC}{+13.9 to +44.8, excess 10.8}%
\newcommand{\txHetOBLFourNeverHinted}{+18.1 to +47.8, excess 8.2}%
\newcommand{\txHetOBLFiveCCplusOlder}{+31.5 to +52.4, excess 7.5}%
\newcommand{\txMaskBaseColorLagOneShare}{29.6\%}%
\newcommand{\txMaskBaseColorLagOneGap}{+55}%
\newcommand{\txMaskLagTwoThreeShareBase}{1.5\%}%
\newcommand{\txMaskLagTwoThreeShareImm}{16.7\%}%
\newcommand{\txMaskLagTwoThreeGapImm}{+53}%
\newcommand{\txMaskLagFourShareBase}{3.1\%}%
\newcommand{\txMaskLagFourShareImm}{6.1\%}%
\newcommand{\txMaskLagFourGapImm}{+32}%
\newcommand{\txMaskPersistColorShare}{0.0\%}%
\newcommand{\txMaskFullyTypedBase}{35.6\%}%
\newcommand{\txMaskFullyTypedPersist}{60.0\%}%
\newcommand{\txMaskOBLTwoDelayedGapBase}{+17}%
\newcommand{\txMaskOBLFiveDelayedGapBase}{+41}%
\newcommand{\txMaskOBLTwoLagTwoThreeShareImm}{2.5\%}%
\newcommand{\hhGamesTwoTier}{425}%
\newcommand{\hhPlaysTwoTier}{9017}%
\newcommand{\hhPlayersTwoTier}{155}%
\newcommand{\hhPairsTwoTier}{95}%
\newcommand{\hhGapJointTwoTier}{+26.16~[+24.05, +28.28]}%
\newcommand{\hhGapJointTwoTierPt}{+26.16}%
\newcommand{\hhGapJointTwoTierCiGame}{[+25.32, +27.01]}%
\newcommand{\hhGapMarginalTwoTierPt}{+28.20}%
\newcommand{\hhGamesTierOne}{598}%
\newcommand{\hhPlaysTierOne}{13089}%
\newcommand{\hhGapJointTierOne}{+27.52~[+25.21, +29.69]}%
\newcommand{\hhGapJointTierOneCiGame}{[+26.79, +28.26]}%
\newcommand{\hhSitFreshColorGap}{+40.9~[+38.2, +43.4]}%
\newcommand{\hhSitFreshColorGapPt}{+40.9}%
\newcommand{\hhSitFreshRankGapPt}{+20.1}%
\newcommand{\hhSitHintedEarlierGapPt}{+16.8}%
\newcommand{\hhSitUnhintedNewestGap}{+48.8~[+38.4, +55.6]}%
\newcommand{\hhSitUnhintedNewestGapPt}{+48.8}%
\newcommand{\hhSitUnhintedOlderGap}{+43.2~[+31.1, +51.5]}%
\newcommand{\hhSitUnhintedOlderGapPt}{+43.2}%
\newcommand{\hhUnhintedShareCombined}{5.4}%
\newcommand{\hhSitOblFiveFreshColorGap}{+40.3}%
\newcommand{\hhSitOblFiveUnhintedNewestGap}{+23.5}%
\newcommand{\hhTransferRsqStrictTree}{0.28}%
\newcommand{\hhTransferRsqPartitionTree}{0.37}%
\newcommand{\hhTransferRsqHumanTree}{0.39}%
\newcommand{\hhTransferRsqFiveSituations}{0.12}%
\newcommand{\hhTransferPearson}{0.81}%
\newcommand{\hhTransferPearsonExclZeroCells}{0.70}%
\newcommand{\hhTransferNumCats}{26}%
\newcommand{\hhTransferZeroCells}{6}%
\newcommand{\hhCatIndirectColorHuman}{+61.3~[+54.4, +66.0]}%
\newcommand{\hhCatIndirectColorHumanPt}{+61.3}%
\newcommand{\hhCatIndirectColorOblFive}{+13.7}%
\newcommand{\hhCatIndirectColorN}{238}%
\newcommand{\hhCatIndirectRankHuman}{+41.1~[+31.9, +48.8]}%
\newcommand{\hhCatIndirectRankHumanPt}{+41.1}%
\newcommand{\hhCatIndirectRankOblFive}{+10.3}%
\newcommand{\tTwelveANeverHintedN}{193}%
\newcommand{\tTwelveANeverHintedShare}{81.1}%
\newcommand{\tTwelveANeverHintedGap}{+63.9}%
\newcommand{\tTwelveAColorOnlyShare}{18.1}%
\newcommand{\tTwelveAColorOnlyGap}{+50.5}%
\newcommand{\hhExpOneGameGap}{+23.6~[+20.2, +26.8]}%
\newcommand{\hhExpOneGamePairs}{36}%
\newcommand{\hhExpTwoToFourGap}{+25.7~[+23.5, +27.8]}%
\newcommand{\hhExpFivePlusGap}{+26.6~[+23.8, +29.7]}%
\newcommand{\hhExpNearStrangersGap}{+22.1~[+17.8, +26.0]}%
\newcommand{\chNonzeroCats}{20}%
\newcommand{\chTaughtCats}{16}%
\newcommand{\chBootDraws}{2000}%
\newcommand{\chOBLTwoCoreRange}{+47 to +55}%
\newcommand{\chHumanCoreRange}{+51--+52}%
\newcommand{\chPbelowFiveCoreRange}{0.89--1.00}%
\newcommand{\chEdgeDrawsMTwo}{1979/2000}%
\newcommand{\chEdgeDrawsMOne}{2000/2000}%
\newcommand{\chBetaMTwo}{+14.3~[+13.2, +15.4]}%
\newcommand{\chWrTwoMTwo}{0.58}%
\newcommand{\chWrTwoMOne}{0.10}%
\newcommand{\chCultureShare}{48\%}%
\newcommand{\chCHShare}{10\%}%
\newcommand{\chDeltaAICc}{-12.6}%
\newcommand{\chWFiveAtTau}{0.48}%
\newcommand{\chSharedTau}{4.27~[3.73, 5.04]}%
\newcommand{\chSharedWrTwo}{0.977}%
\newcommand{\chSharedPedge}{0.00}%
\newcommand{\chSharedPinOneThree}{0.00}%
\newcommand{\chSharedWOne}{8\%}%
\newcommand{\chSharedNCells}{4}%
\newcommand{\tOneFourContribSingleOne}{7.1}%
\newcommand{\tOneFourCostSingleOne}{-0.39~[-0.51, -0.26]}%
\newcommand{\tOneFourCostSingleOnePt}{-0.39}%
\newcommand{\tOneFourExcessSingleOne}{-0.12~[-0.25, +0.01]}%
\newcommand{\tOneFourExcessSingleOnePt}{-0.12}%
\newcommand{\tOneFourExcessCISingleOne}{[-0.25, +0.01]}%
\newcommand{\tOneFourTotalGapSingleOnePt}{18.9}%
\newcommand{\tOneFourContribSingleTwo}{5.3}%
\newcommand{\tOneFourCostSingleTwoPt}{-0.36}%
\newcommand{\tOneFourExcessSingleTwo}{-0.09~[-0.23, +0.03]}%
\newcommand{\tOneFourExcessSingleTwoPt}{-0.09}%
\newcommand{\tOneFourExcessCISingleTwo}{[-0.23, +0.03]}%
\newcommand{\tOneFourTotalGapSingleTwoPt}{19.0}%
\newcommand{\tOneFourContribSingleThree}{1.2}%
\newcommand{\tOneFourCostSingleThreePt}{-0.07}%
\newcommand{\tOneFourExcessSingleThreePt}{-0.02}%
\newcommand{\tOneFourExcessCISingleThree}{[-0.12, +0.09]}%
\newcommand{\tOneFourTotalGapSingleThreePt}{21.1}%
\newcommand{\tOneFourContribSingleFour}{1.1}%
\newcommand{\tOneFourCostSingleFourPt}{-0.01}%
\newcommand{\tOneFourExcessSingleFourPt}{+0.03}%
\newcommand{\tOneFourExcessCISingleFour}{[-0.05, +0.12]}%
\newcommand{\tOneFourTotalGapSingleFourPt}{21.1}%
\newcommand{\tOneFourContribSingleFive}{1.0}%
\newcommand{\tOneFourCostSingleFivePt}{-0.01}%
\newcommand{\tOneFourExcessSingleFivePt}{+0.02}%
\newcommand{\tOneFourExcessCISingleFive}{[-0.04, +0.09]}%
\newcommand{\tOneFourTotalGapSingleFivePt}{20.9}%
\newcommand{\tOneFourContribSingleSix}{0.6}%
\newcommand{\tOneFourCostSingleSixPt}{+0.10}%
\newcommand{\tOneFourExcessSingleSix}{+0.25~[+0.12, +0.38]}%
\newcommand{\tOneFourExcessSingleSixPt}{+0.25}%
\newcommand{\tOneFourExcessCISingleSix}{[+0.12, +0.38]}%
\newcommand{\tOneFourTotalGapSingleSixPt}{20.7}%
\newcommand{\tOneFourContribSingleSeven}{0.5}%
\newcommand{\tOneFourCostSingleSevenPt}{-0.01}%
\newcommand{\tOneFourExcessSingleSevenPt}{+0.01}%
\newcommand{\tOneFourExcessCISingleSeven}{[-0.06, +0.09]}%
\newcommand{\tOneFourTotalGapSingleSevenPt}{21.0}%
\newcommand{\tOneFourContribSingleEight}{0.5}%
\newcommand{\tOneFourCostSingleEightPt}{-0.02}%
\newcommand{\tOneFourExcessSingleEightPt}{-0.01}%
\newcommand{\tOneFourExcessCISingleEight}{[-0.07, +0.04]}%
\newcommand{\tOneFourTotalGapSingleEightPt}{21.0}%
\newcommand{\tOneFourContribCtrlA}{0.0}%
\newcommand{\tOneFourCostCtrlAPt}{-0.03}%
\newcommand{\tOneFourExcessCtrlA}{+0.05~[-0.08, +0.18]}%
\newcommand{\tOneFourExcessCtrlAPt}{+0.05}%
\newcommand{\tOneFourExcessCICtrlA}{[-0.08, +0.18]}%
\newcommand{\tOneFourTotalGapCtrlAPt}{21.1}%
\newcommand{\tOneFourContribCtrlB}{0.0}%
\newcommand{\tOneFourCostCtrlBPt}{+0.01}%
\newcommand{\tOneFourExcessCtrlB}{+0.09~[+0.01, +0.18]}%
\newcommand{\tOneFourExcessCtrlBPt}{+0.09}%
\newcommand{\tOneFourExcessCICtrlB}{[+0.01, +0.18]}%
\newcommand{\tOneFourTotalGapCtrlBPt}{21.2}%
\newcommand{\tOneFourContribCtrlRlagFour}{0.2}%
\newcommand{\tOneFourCostCtrlRlagFourPt}{-0.07}%
\newcommand{\tOneFourExcessCtrlRlagFour}{-0.00~[-0.10, +0.10]}%
\newcommand{\tOneFourExcessCtrlRlagFourPt}{-0.00}%
\newcommand{\tOneFourExcessCICtrlRlagFour}{[-0.10, +0.10]}%
\newcommand{\tOneFourTotalGapCtrlRlagFourPt}{21.3}%
\newcommand{\tOneFourContribCumTwo}{12.4}%
\newcommand{\tOneFourCostCumTwoPt}{-1.11}%
\newcommand{\tOneFourExcessCumTwoPt}{-0.52}%
\newcommand{\tOneFourExcessCICumTwo}{[-0.74, -0.29]}%
\newcommand{\tOneFourTotalGapCumTwoPt}{15.9}%
\newcommand{\tOneFourContribCumFour}{14.7}%
\newcommand{\tOneFourCostCumFourPt}{-1.27}%
\newcommand{\tOneFourExcessCumFourPt}{-0.55}%
\newcommand{\tOneFourExcessCICumFour}{[-0.81, -0.29]}%
\newcommand{\tOneFourTotalGapCumFourPt}{15.9}%
\newcommand{\tOneFourContribCumSix}{16.3}%
\newcommand{\tOneFourCostCumSixPt}{-1.71}%
\newcommand{\tOneFourExcessCumSixPt}{-0.73}%
\newcommand{\tOneFourExcessCICumSix}{[-1.08, -0.37]}%
\newcommand{\tOneFourTotalGapCumSixPt}{14.7}%
\newcommand{\tOneFourContribCumEight}{17.3}%
\newcommand{\tOneFourCostCumEightPt}{-1.84}%
\newcommand{\tOneFourExcessCumEightPt}{-0.81}%
\newcommand{\tOneFourExcessCICumEight}{[-1.19, -0.44]}%
\newcommand{\tOneFourTotalGapCumEight}{14.3~[+14.00, +14.68]}%
\newcommand{\tOneFourTotalGapCumEightPt}{14.3}%
\newcommand{\eFiveControlRates}{3.76 / 0.61 / 0.24}%
\newcommand{\eFiveBaselineGap}{21.3}%
\newcommand{\eFiveTopCellSharePct}{12.6}%
\newcommand{\eFiveTopCellGap}{+57}%
\newcommand{\eFiveTopCellRandomCost}{-0.27}%
\newcommand{\eFiveSlopeContrib}{-2.87~[-4.78, -0.98]}%
\newcommand{\eFiveRsqContrib}{0.44}%
\newcommand{\eFiveRsqGap}{0.72}%
\newcommand{\eFiveRsqShare}{0.11}%
\newcommand{\eFiveCumTopEightCost}{-1.84}%
\newcommand{\eFiveSumSinglesTopEight}{-0.77}%
\newcommand{\eFiveSuperAddTopEight}{-1.08~[-1.47, -0.72]}%
\newcommand{\eFiveConlyLagTwoThreeShareBase}{1.5}%
\newcommand{\eFiveConlyLagTwoThreeShareTopOne}{5.9}%
\newcommand{\eFiveConlyLagTwoThreeGapTopOne}{+54.2}%
\newcommand{\eFiveConlyLagTwoThreeShareTopEight}{16.3}%
\newcommand{\eFiveConlyLagTwoThreeGapTopEight}{+52.9}%
\newcommand{\eFivebExcessOBLFiveA}{+0.14~[-0.003, +0.292]}%
\newcommand{\eFivebExcessOBLFiveB}{+0.15~[+0.035, +0.274]}%
\newcommand{\eFivebExcessOBLFourA}{+0.06~[-0.098, +0.205]}%
\newcommand{\eFivebPooledExcess}{+0.12~[+0.04, +0.20]}%
\newcommand{\eFivebDecision}{INCONCLUSIVE}%
\newcommand{\eFivebNumAgentCIsExcludeZero}{1}%
\newcommand{\eFivebRankHintSubstitutionRange}{48--62\%}%
\newcommand{\eFiveGateLogOnlyIdentical}{1000/1000}%
\newcommand{\eFiveGateDescriptorAgreement}{24838/24838}%
\newcommand{\eFiveDeckIdentity}{1000/1000 for every condition}%
\newcommand{\tFiveShareFreshColorOBLOne}{18.1}%
\newcommand{\tFiveShareFreshRankOBLOne}{36.9}%
\newcommand{\tFiveGapFreshRankOBLOnePt}{+3.7}%
\newcommand{\tFiveShareStaleHintedOBLOne}{40.6}%
\newcommand{\tFiveGapStaleHintedOBLOnePt}{+1.3}%
\newcommand{\tFiveShareUnhintedNewestOBLOne}{2.0}%
\newcommand{\tFiveGapUnhintedNewestOBLOne}{-13.9~[-17.6, -9.9]}%
\newcommand{\tFiveGapUnhintedNewestOBLOnePt}{-13.9}%
\newcommand{\tFiveShareUnhintedOtherOBLOne}{2.3}%
\newcommand{\tFiveGapUnhintedOtherOBLOnePt}{-9.8}%
\newcommand{\tFiveGapFreshColorOBLOneTrend}{+1.0}%
\newcommand{\tcOpeningGapOBLTwo}{+3.5}%
\newcommand{\tcEndgameGapOBLTwo}{+4.1}%
\newcommand{\tcPeakGapOBLFive}{+36.7}%
\newcommand{\tcPeakTurnBinOBLFive}{20}%
\newcommand{\tcOpeningGapOBLFive}{+6.8}%
\newcommand{\tcEndgameGapOBLFive}{+6.2}%
\newcommand{\tFiveShareFreshColorOBLFive}{40.4}%
\newcommand{\tFivePostFreshColorOBLFive}{41.1}%
\newcommand{\tFiveLossFreshColorOBLFive}{0.8}%
\newcommand{\tFiveGapFreshColorOBLFive}{+40.3~[+39.7, +40.9]}%
\newcommand{\tFiveShareFreshRankOBLFive}{24.3}%
\newcommand{\tFiveGapFreshRankOBLFivePt}{+11.9}%
\newcommand{\tFiveShareStaleHintedOBLFive}{33.7}%
\newcommand{\tFiveGapStaleHintedOBLFivePt}{+6.4}%
\newcommand{\tFiveShareUnhintedNewestOBLFive}{0.8}%
\newcommand{\tFiveGapUnhintedNewestOBLFive}{+23.5~[+17.3, +29.6]}%
\newcommand{\tFiveGapUnhintedNewestOBLFivePt}{+23.5}%
\newcommand{\tFiveShareUnhintedOtherOBLFive}{0.7}%
\newcommand{\tFiveGapUnhintedOtherOBLFivePt}{+18.3}%
\newcommand{\tFiveGapFreshColorOBLFiveTrend}{+40.3}%
\newcommand{\tcOBLOneRangeAllBins}{+0.3..+4.6}%
\newcommand{\tcFreshColorShareOfGapOBLFive}{75.1}%
\newcommand{\sSixShareOneHintOBLOne}{39}%
\newcommand{\sSixShareTwoPlusOBLOne}{56}%
\newcommand{\sSixPostOneHintOBLOne}{12.8}%
\newcommand{\sSixLossOneHintOBLOne}{7.8}%
\newcommand{\sSixMeanPostOBLOne}{7.8}%
\newcommand{\sSixShareOneHintOBLFive}{61}%
\newcommand{\sSixShareTwoPlusOBLFive}{38}%
\newcommand{\sSixPostOneHintOBLFive}{37.3}%
\newcommand{\sSixLossOneHintOBLFive}{2.9}%
\newcommand{\sSixMeanPostOBLFive}{24.1}%
\newcommand{\sSixPosteriorRiseCompositionPP}{+1.0}%
\newcommand{\sSixPosteriorRiseWithinPP}{+15.2}%
\newcommand{\tSixFreshGapOBLOnexOBLTwoLowPt}{+1.3}%
\newcommand{\tSixFreshShareOBLOnexOBLTwoHigh}{21}%
\newcommand{\tSixFreshLossOBLOnexOBLTwoHigh}{4.6}%
\newcommand{\tSixFreshGapOBLOnexOBLFourLowPt}{+2.0}%
\newcommand{\tSixFreshGapOBLOnexOBLFiveLow}{+1.9~[+1.3, +2.4]}%
\newcommand{\tSixFreshShareOBLOnexOBLFiveHigh}{27}%
\newcommand{\tSixFreshLossOBLOnexOBLFiveHigh}{13.3}%
\newcommand{\tSixFreshGapOBLOnexOBLFiveHighPt}{+14.1}%
\newcommand{\tSixFreshLossOBLTwoxOBLFiveLow}{0.6}%
\newcommand{\tSixFreshGapOBLTwoxOBLFiveLowPt}{+30.6}%
\newcommand{\tSixFreshGapOBLTwoxOBLFiveHighPt}{+23.1}%
\newcommand{\tNineSelfplayFreshLossRangeLTwotoFive}{0.8..1.0}%
\newcommand{\tNineOBLOneFreshLossRangeWithHigher}{0.7..1.4}%
\newcommand{\axGapJoint}{+27.62~[+27.27, +27.97]}%
\newcommand{\axNGames}{1752}%
\newcommand{\axGapJointFullOneEightFiveEight}{+27.59~[+27.26, +27.92]}%
\newcommand{\axOursGapJointGameCl}{+26.16~[+25.31, +26.99]}%
\newcommand{\axMeanScore}{23.45}%
\newcommand{\axRTwoStrictOblFiveTree}{0.338}%
\newcommand{\axPearsonVsOursExclZero}{0.97}%
\newcommand{\axChanIndirectColor}{+48.9~[+45.4, +52.1]}%
\newcommand{\axChanIndirectColorPt}{+48.9}%
\newcommand{\axChanIndirectRank}{+39.9~[+36.2, +43.5]}%
\newcommand{\axChanIndirectRankPt}{+39.9}%
\newcommand{\axChanUnhintedNewest}{+37.2~[+31.9, +42.2]}%
\newcommand{\axSitFreshColorGap}{+42.7~[+42.2, +43.3]}%
\newcommand{\axSitFreshColorShare}{31.5}%
\newcommand{\axSitFreshRankGap}{+22.1~[+21.5, +22.7]}%
\newcommand{\axSitFreshRankShare}{21.6}%
\newcommand{\axSitStaleHintedGap}{+19.4~[+18.9, +19.9]}%
\newcommand{\axSitStaleHintedShare}{45.2}%
\newcommand{\axSitUnhintedNewestGap}{+37.2~[+31.9, +42.2]}%
\newcommand{\axSitUnhintedNewestShare}{1.1}%
\newcommand{\axSitUnhintedOtherGap}{+35.3~[+28.0, +41.9]}%
\newcommand{\axSitUnhintedOtherShare}{0.7}%
\newcommand{\axUnhintedShareTotal}{1.7}%
\newcommand{\bcHCZeroScoreRateSelf}{100.0}%
\newcommand{\bcHCChanIndirectColorXObl}{-5.7~[-21.6, +12.9]}%
\newcommand{\bcHCChanIndirectRankXObl}{-8.1~[-19.0, +3.9]}%
\newcommand{\bcHCChanUnhintedNewestXObl}{-32.0~[-37.6, -24.4]}%
\newcommand{\bcHCChanUnhintedOtherXObl}{-17.6~[-24.5, -9.6]}%
\newcommand{\bcHCFreshColorNewestThreepXObl}{+59.9~[+58.0, +61.7]}%
\newcommand{\bcOCMatchedZeroScoreRateSelf}{97.2}%
\newcommand{\bcOCMatchedProfileRVsOblXObl}{0.929}%
\newcommand{\bcOCMatchedChanUnhintedNewestXObl}{+2.6~[-11.2, +16.7]}%
\newcommand{\bcOCTwentyKGapSelf}{+21.22~[+20.87, +21.58]}%
\newcommand{\bcOCTwentyKProfileRVsOblSelf}{0.967}%
\newcommand{\bcOCTwentyKFreshColorNewestThreepSelf}{+56.9~[+56.2, +57.6]}%
\newcommand{\bcOCScalingR}{0.842 / 0.916 / 0.967}%
\newcommand{\hhMeanScore}{18.52}%
\newcommand{\hhSitFreshRankShare}{24.5}%
\newcommand{\hhSitStaleHintedShare}{41.1}%
\newcommand{\hhSitFreshColorShare}{29.0}%
\newcommand{\hhSitUnhintedNewestShare}{2.9}%
\newcommand{\hhSitUnhintedOtherShare}{2.5}%
\newcommand{\hhCatIndirectColorShare}{2.64}%
\newcommand{\hhCatIndirectRankShare}{1.33}%
\newcommand{\oblFiveCatIndirectColorShare}{0.28}%
\newcommand{\oblFiveCatIndirectRankShare}{0.19}%
\newcommand{\oblOneCatIndirectColorShare}{0.40}%
\newcommand{\oblOneCatIndirectRankShare}{0.92}%
\newcommand{\axCatIndirectColorShare}{1.08}%
\newcommand{\axCatIndirectRankShare}{0.81}%
\newcommand{\hhScoreFloorGap}{+27.20~[+26.29, +28.04]}%
\newcommand{\hhScoreFloorGames}{338}%
\newcommand{\sharedTaxCells}{21}%
\newcommand{\sharedTaxAggHuman}{+25.81}%
\newcommand{\sharedTaxAggOblFive}{+21.12}%
\newcommand{\sharedTaxRate}{+5.73}%
\newcommand{\sharedTaxMix}{-1.03}%
\newcommand{\tcfigPeakGapOBLTwo}{+21.7~[+20.3, +23.1]}%
\newcommand{\tcfigPeakGapOBLFive}{+36.7~[+35.2, +38.2]}%
\newcommand{\tcfigOblFiveLossRateRange}{1.9-3.2}%
\newcommand{\tcfigHumanPeakGap}{+39.2}%
\newcommand{\tcfigHumanOpeningGap}{+18.3}%
\newcommand{\tcfigHumanLateBinsAboveOblFive}{6/6}%
\newcommand{\tcfigPartnerPeakOblFiveSelf}{+36.7}%
\newcommand{\tcfigPartnerPeakOblFiveWithOblThree}{+30.7}%
\newcommand{\tcfigPartnerPeakOblFiveWithOblOne}{+13.8}%
\newcommand{\tcfigMaskPeakOblFiveBaseline}{+36.6}%
\newcommand{\tcfigMaskPeakOblFiveImmediacy}{+26.4}%
\newcommand{\tcfigMaskPeakOblFivePersistent}{+9.4}%
\newcommand{\axColorKnownSubcellGap}{+50.4~[+47.1, +53.6]}%
\newcommand{\axColorKnownSubcellGapPt}{+50.4}%
\newcommand{\bcHCChanUnhintedNewestLossXObl}{95.5}%
\newcommand{\chBetaPairCl}{+12.0~[+10.5, +13.3]}%
\newcommand{\chCultureSharePairCl}{34}%
\newcommand{\axNGamesReleased}{1858}%
\newcommand{\axMinusOursGameCl}{+1.46}%
\newcommand{\nMinThreeTargetOBLFive}{386}%
\newcommand{\axDroppedStuckGames}{106}%
\newcommand{\axScoreFloorMin}{13}%
\newcommand{\chWrTwoMOnePairCl}{0.41}%
\newcommand{\chWrTwoMTwoPairCl}{0.74}%
\newcommand{\fcColOblFiveUnhintedGapPt}{+6.1}%
\newcommand{\fcColOblFiveColorOnlyGapPt}{+26.3}%
\newcommand{\fcRnkHumanUnhintedGapPt}{+45.4}%
\newcommand{\fcRnkOblFiveUnhintedGapPt}{+1.6}%
\newcommand{\ldDAllOne}{19.0~[+16.9, +20.8]}%
\newcommand{\ldDAllTwo}{10.2~[+8.4, +11.9]}%
\newcommand{\ldDAllThree}{8.3~[+6.6, +10.0]}%
\newcommand{\ldDAllFour}{7.1~[+6.1, +8.9]}%
\newcommand{\ldDAllFive}{7.0~[+5.9, +8.7]}%
\newcommand{\ldDAllTwoMinusFive}{+3.2~[+1.7, +3.7]}%
\newcommand{\ldDCoreOne}{15.9~[+13.9, +18.0]}%
\newcommand{\ldDCoreOnePt}{15.9}%
\newcommand{\ldDCoreTwo}{4.1~[+2.6, +5.9]}%
\newcommand{\ldDCoreTwoPt}{4.1}%
\newcommand{\ldDCoreThree}{2.7~[+1.5, +4.5]}%
\newcommand{\ldDCoreFour}{1.3~[+1.3, +3.5]}%
\newcommand{\ldDCoreFive}{1.7~[+1.5, +3.7]}%
\newcommand{\ldDCoreOneMinusTwo}{+11.9~[+10.4, +13.0]}%
\newcommand{\ldDCoreTwoMinusFive}{+2.4~[-0.6, +3.5]}%
\newcommand{\ldNSharedCells}{21}%
\newcommand{\ldNExcludedChannelCells}{11}%
\newcommand{\silHHAllAllN}{484}%
\newcommand{\silHHAllAllGap}{+46.2~[+35.9, +52.6]}%
\newcommand{\silHHAllHintOtherN}{286}%
\newcommand{\silHHAllHintOtherGap}{+57.9~[+50.6, +62.7]}%
\newcommand{\silHHAllColorOtherN}{193}%
\newcommand{\silHHAllColorOtherPost}{69.6}%
\newcommand{\silHHAllColorOtherLoss}{5.7}%
\newcommand{\silHHAllCleanN}{198}%
\newcommand{\silHHAllCleanGap}{+29.4~[+17.3, +40.1]}%
\newcommand{\silAXAllAllN}{736}%
\newcommand{\silAXAllAllGap}{+36.5~[+31.5, +40.9]}%
\newcommand{\silAXAllHintOtherN}{229}%
\newcommand{\silAXAllHintOtherGap}{+43.0~[+35.2, +49.4]}%
\newcommand{\silAXAllCleanN}{507}%
\newcommand{\silAXAllCleanGap}{+33.5~[+27.7, +39.1]}%
\newcommand{\silOBLFiveAllAllN}{394}%
\newcommand{\silOBLFiveAllAllGap}{+21.0~[+16.8, +25.4]}%
\newcommand{\silOBLFiveAllHintOtherN}{55}%
\newcommand{\silOBLFiveAllHintOtherGap}{+2.5~[-9.0, +14.2]}%
\newcommand{\silOBLFiveAllCleanN}{339}%
\newcommand{\silOBLFiveAllCleanGap}{+24.0~[+19.4, +28.7]}%
\newcommand{\silOBLFiveAllCleanGapPt}{+24.0}%
\newcommand{\silDiffHHMinusOBLFiveAllClean}{+5.3~[-7.7, +17.1]}%
\newcommand{\silDiffAXMinusOBLFiveAllClean}{+9.5~[+2.0, +16.8]}%
\newcommand{\chTaughtCatsPairCl}{15}%

\newcommand{\hhSitHintedEarlierShare}{\hhSitStaleHintedShare}%

%% file: sections/00_abstract.tex
Cooperative success rests in part on implicit communication. Partners converge on
conventions: regularities that give an action meaning beyond its direct effect. Yet
conventions are usually observed only indirectly: a score drop when partners are swapped
shows that a convention existed, but not what it was or where it was used. In the card
game Hanabi, the exact probability that a play fails can be computed from the hints
received and the cards in view. Players who use conventions succeed more often than this
prediction allows; the resulting margin is the convention gap. Off-belief learning provides a
hierarchy of agents (levels 1--5) whose first level uses no conventions by construction
and whose later levels use progressively more. For humans and these agents on one scale,
we ask what the conventions are and where they are used. On what: in two independently
collected corpora of human games, humans show a larger gap than every agent level. On
the conventions both populations use, however, humans do no better than level 2. The
human advantage is concentrated in a situation that no agent level learns to use:
hints that point to a card other than the ones they touch. On where: conventions depend on what the player already knows about the card, not
on how recently it was hinted, and they are used most in the mid-game; humans use them
earlier and later in the game than the agents. In interventions on the agents, blocking
a single high-gap situation costs little, because meaning reroutes to other situations,
but blocking several together costs far more than the sum of the single costs. Thus, the
convention gap turns conventions from an inferred cause of score loss into a measurable,
comparable, and causally testable object.

%% file: sections_aamas/01_intro.tex
\section{Introduction}
\label{sec:intro}

Successful cooperation depends in part on implicit communication. Partners converge on
conventions: regularities that give an action a meaning beyond its direct effect
\citep{lewis1969}. People and AI agents both rely on conventions alongside the explicit
information the rules provide. In zero-shot coordination (ZSC) \citep{hu2020otherplay,treutlein2021zsc}, conventions are what
break when a policy meets a partner it was not trained with. Yet conventions are
usually studied indirectly, through the score an agent loses when its partner is
swapped. A score drop shows that a convention existed. It does not say what the
convention was, where it operated, what it was worth, or whether humans share it.

A recent study \citep{paper1} proposes a more direct diagnostic in the cooperative card game Hanabi \citep{bard2020hanabi}. At every play action, a
literal observer computes the exact probability that the played card fails, using only
information the rules guarantee: the hint constraints on every card in the hand, and
card counting over all visible zones. Averaged over many plays, the difference between
this predicted failure rate and the actual one is the \emph{convention gap}
\metriccite{}. Off-belief learning (OBL) trains a hierarchy of agents \citep{hu2021obl}.
Level~1 is trained to read past actions only for their literal content, and later
levels may interpret more. On this hierarchy, the gap rose monotonically with level,
from $\gapOBLOnePt$ to $\gapOBLFivePt$ percentage points (pp) \paperone{}.

Here, we identify candidate conventions where plays beat the literal
prediction, and quantify that departure in pp. We ask three questions of humans and the OBL hierarchy on one scale. (i)~\emph{What} are the conventions: which situations carry the
gap, and to what extent do the two populations agree on them? (ii)~\emph{Where} are they used:
when in the game, in which situations, and carried by which feature of the game
state? (iii)~To what extent does removing a convention cost score, relative to
removing an equally frequent play at random? We address them
by breaking the gap down by situation and by turn, by placing a public human corpus \citep{hanablive} and a second, independently collected corpus \citep{ah2ac2} on the same scale as the hierarchy, and by intervening on the
agents' policies against a frequency-matched control.

\paragraph{Findings in brief.}
The convention gap of the human corpus, $\hhGapJointTwoTier\pp$, is larger than
that of every level of the OBL hierarchy, and their profile across play situations replicates on the second corpus ($r = \axPearsonVsOursExclZero$; Appendix~\ref{app:corpus}). On the conventions both
populations share, humans are not significantly closer to level~5 than to
level~2. The
remaining difference lies in one situation where no level approaches the human gap: hints that point to a card other than the one they touch (\emph{indirect reference}; \S\ref{sec:what}). The gap peaks in
the mid-game, and the human gap exceeds level~5's in the opening and late game. In the agents, the gap follows what the player already knows about the played card rather than how recently it was hinted (\S\ref{sec:when}). Finally,
blocking one high-gap situation in the agents' play lowers their score little more
than blocking an equally frequent play at random, because blocked plays partly reappear
in other situations; blocking several situations together costs more than
the sum of the single blocks ($\eFiveCumTopEightCost$ points jointly against
$\eFiveSumSinglesTopEight$ summed; \S\ref{sec:worth}). Together, these results show how
an exact literal baseline can locate and compare conventions across humans and
agents, and price them in agents.

%% file: sections/01b_related.tex
\subsection{Related work}
\label{sec:related}

\paragraph{Conventions and zero-shot coordination.}
A convention in Lewis's sense is a self-sustaining regularity that resolves a
coordination problem among alternatives \citep{lewis1969}. In cooperative multi-agent
reinforcement learning such regularities are what self-play policies acquire, and the
Hanabi challenge made their cost visible: policies that score well together fail with
partners trained separately \citep{bard2020hanabi}. The ZSC
program addresses this by construction. Other-Play removes symmetry-dependent
conventions \citep{hu2020otherplay}; off-belief learning trains its first
level to read past actions only for their literal content and lets later levels interpret more \citep{hu2021obl}; K-level reasoning,
trajectory diversity, and the label-free formalism of \citet{treutlein2021zsc} give
further routes to compatible policies \citep{cui2021klevel,lupu2021trajedi}. Evaluation
in this line is by cross-play score \citep{wolski2026crossplay}. This paper proposes no
ZSC method; it locates the conventions those methods act on, using the
off-belief learning hierarchy.

\paragraph{Human--AI coordination and human proxies.}
Evaluating agents with people is costly, so the field increasingly substitutes
behavioral clones of human play. The practice was established in Overcooked
\citep{carroll2019utility}; population-based training aims to remove the need for human
data altogether \citep{strouse2021fcp}; in Hanabi, human-team studies
\citep{siu2021hanabi} and the AH2AC2 benchmark, whose proxies are cloned from human
games \citep{ah2ac2}, follow the same route. Appendix~\ref{app:clones} tests what such clones preserve at the level of individual
conventions.

\paragraph{Reading conventions out of policies.}
\citet{shih2021conventions} separate a policy into a task-specific module and a
partner-specific convention module and show that the latter transfers across tasks;
Any-Play augments training to discourage conventions \citep{lucas2022anyplay}. Both
treat conventions as a hidden variable to be separated out or suppressed. Here plays
are compared with an exact literal baseline, so a convention's location and carrier
can be reported without a learned probe.

\paragraph{Implicit communication and pragmatics.}
Hints in Hanabi are speech acts whose intended meaning exceeds their literal content, the
situation the rational speech act framework models as recursive inference between
speaker and listener \citep{frank2012rsa,goodman2016rsa}, building on Gricean implicature
and common ground \citep{grice1975,clark1996}. The convention gap needs no fitted listener:
it compares observed failures with literal predictions, in percentage points of failure.

\paragraph{Depth of reasoning.}
Level-$k$ and cognitive-hierarchy models estimate how many steps of recursive reasoning
people use, and typically find one to three \citep{stahl1995,nagel1995,camerer2004ch}.
Appendix~\ref{app:depth} asks whether the human--agent difference is a
difference of depth in that sense, and finds that depth alone does not account for it.

%% file: sections_aamas/02_setup.tex
\section{Methods}
\label{sec:setup}

\subsection{The convention gap}
\label{sec:metric}
We use the metric as defined in a previous study \citep{paper1}. At every play action, an exact
literal observer computes $P(\text{life loss})$ from rule-guaranteed information only: the positive and negative hint constraints on every
card in the hand, and card counting over all visible zones. Because a hint on one card
also constrains the others, the observer uses the joint distribution over the whole
hand rather than per-card marginals. For a set of plays,
\begin{equation}
  \Delta \;=\; \operatorname{mean}\big(P(\text{life loss})\big) \;-\; \text{actual life-loss rate},
  \label{eq:gap}
\end{equation}
reported in percentage points. A positive value means plays fail less often than
literal information predicts. The gap is signed and cannot exceed the mean
$P(\text{life loss})$. It is exactly zero on plays whose outcome literal
information fixes, and subsets with gaps of opposite sign can cancel. A value near
zero therefore does not show that partners conveyed nothing beyond literal content. By
quantifying conventions we mean locating where plays depart from this literal
prediction and by how many percentage points. That a positive gap reflects conventions is an
interpretation, not a definition: a partner who hints playable cards also
produces one. The interpretation is consistent with the level ordering of
\S\ref{sec:hierarchy} and the decompositions of \S\ref{sec:what}--\S\ref{sec:worth}.

\paragraph{Statistics.}
Each outcome is $0$ or $1$, so $\Delta$ is always a mean over many plays. Plays within
a game share a deck and a history, so all intervals resample whole games (whole pairs,
for humans) rather than individual plays \citep{cameron2008cluster}. Draw counts are
fixed in advance. Where many data-induced categories (\S\ref{sec:where}) are tested we control the
false discovery rate at $q = 0.05$ per level \citep{benjamini1995fdr}. Wherever
categories are induced from data, they are defined on a discovery half and every
reported statistic is computed on a held-out estimation half, so no pattern is scored
on the data that found it. Analyses with pre-specified categories use all plays.

\subsection{The OBL hierarchy}
\label{sec:hierarchy}
Off-belief learning \citep{hu2021obl} trains a policy whose $Q$-targets are computed on
fictitious states, resampled from a belief model that assumes all past actions came
from a fixed base policy $\pi_0$. The learner is scored as if the past carried only
what $\pi_0$ would put into it. Level~1 uses a uniformly random $\pi_0$, so it is
trained to read past actions only for their literal content (\emph{grounded} play).
Each later level is trained with beliefs induced by the previous level's policy, so it may interpret more. Agent levels are written level~$k$ in the text and OBL$k$ in figures and tables. The released
archive holds five seed models per level; the ladder below uses
the first of each. Level~5 is in the archive but not among the results published with OBL \citep{hu2021obl},
and every trend claim here holds on levels 1--4 alone. In the previous study \citep{paper1}, the
gap rose monotonically with level, from $\gapOBLOne\pp$ at level~1 to
$\gapOBLFive\pp$ at level~5. Level-1 cross-play, pooled over all pairings among five
independently trained level-1 models, gave $\gapOBLOneCross\pp$.

\paragraph{Agent data.}
Unless stated otherwise, agent conditions use $1000$ games per condition with distinct
environment seeds and greedy acting; the taxonomy below pools five seed models per
level ($5000$ games; level~1 uses only the first).

\subsection{Human data}
\label{sec:corpus}

People who play Hanabi competitively have conventions too, but theirs are
\emph{taught}: a player community publishes and teaches a convention system \citep{hgroup}. The human games are public logs from the hanab.live platform \citep{hanablive}, where
this community plays. We first establish that these logs and the generated agent games
are comparable. First, the same exact
joint posterior is applied to both populations; nothing is re-estimated for humans.
Second, the human games were replayed through the \emph{same} extraction pipeline used
for the agents and reproduced the reference posteriors bit-identically over all
$\hhPlaysTwoTier$ plays. Third, the corpus
and its filters are inherited unchanged from the previous study \citep{paper1}: a
completed-games filter and a two-tier bot exclusion applied by the same code, giving
$\hhGamesTwoTier$ games, $\hhPlaysTwoTier$ plays and $\hhPlayersTwoTier$ players. Pairs
recur across games ($\hhPairsTwoTier$ pairs in all), so every human interval is
clustered at the \emph{pair} level. A second, independently collected corpus, the open
games of the AH2AC2 benchmark \citep{ah2ac2}, is processed through the identical
pipeline and serves as a replication corpus.

\subsection{Situations and the induced taxonomy}
\label{sec:where}

Because $\Delta$ is a mean over plays, it can be broken down by situation.
Two category systems do this. The first is hand-crafted: five mutually exclusive play
situations (played immediately after a partner color hint touched the card;
immediately after a rank hint; hinted earlier; unhinted and newest-drawn; unhinted and older). A card is unhinted if no hint has touched it; hints to other cards can still exclude values for it. Table~\ref{tab:usage} and \S\ref{sec:when} report the gap by situation. Plays made immediately after a hint touched a \emph{different} card,
\emph{indirect reference}, are tracked with the finer categories described next.

The second system is induced from the data. Every play is described by six components
of its context: the partner's
immediately preceding action, its relation to the board, the card's ordered hint
history, the lag since the last hint that touched it, hand position, and the
hint-token stock. Games are split 50/50 into discovery and estimation halves. Every
context type that clears a minimum cell size $n_{\min}$ on the discovery half becomes
its own category, and rarer types merge by hierarchical backoff, so there is no opaque
\emph{other} bucket. $n_{\min}$ is derived from the empirical residual standard
deviation at a $\pm 5\pp$ precision target, giving \txNminRange{} across levels. A
depth-4 CART tree \citep{breiman1984cart} fit on the same discovery half gives a
compact summary of each taxonomy (Appendix~\ref{app:metric}). All reported statistics
come from the estimation half.  The level-5
categories, once fit, are also kept fixed and applied unchanged to other populations;
we call this the \emph{frozen} level-5 taxonomy.

For comparing populations, a \emph{shared} taxonomy is defined jointly on all six: a
cell exists only if every population's discovery half supports it at its own
$n_{\min}$, so every value can be estimated for everyone. Each cell is labeled by the
components that define it; Fig.~\ref{fig:shared} gives the key. We call any situation or category with a positive gap a \emph{channel}.

\subsection{Interventions}
\label{sec:maskdesign}

Categories are observational: a category's gap says how far its plays depart from the
literal prediction, not what would happen if the channel were removed. We therefore
intervene on the agents' policies. When the acting player's greedy action is to play a
card whose context falls in a target category, that play is removed from the legal
set and the best remaining action is taken from the same network outputs. The network
is not re-run, and its recurrent state follows the real observations. All conditions
share $1000$ environment seeds, so scores are compared per deck. Verification gates
precede every full run (Appendix~\ref{app:interventions}). Every score difference
reported in \S\ref{sec:worth} is therefore between the same decks and the same network.

The key part of the design is its control. Blocking \emph{any} frequent play disturbs
the game whatever its convention content, so a raw score drop says nothing about
conventions. We therefore run \emph{frequency-matched random masks}: the greedy play
is blocked with a probability chosen to match a target category's trigger rate, with
no regard to its content, and these masks define a null cost curve. The quantity of
interest is the \emph{excess cost}: the paired score cost minus the random-mask cost
at the same trigger rate, with the control curve re-fit inside every bootstrap draw.
Three low-gap control categories at level~5 (two fully-identified-card cells and the large lag-$4{+}$ rank cell) check the control curve (\S\ref{sec:worth}). Negative score changes are losses.

%% file: sections_aamas/03_what.tex
\section{Results}
\label{sec:results}

\subsection{What conventions are used: humans and agents on one scale}
\label{sec:what}\label{sec:humans}

\begin{figure*}[tp]
  \centering
  \includegraphics[width=\textwidth,height=0.78\textheight,keepaspectratio]{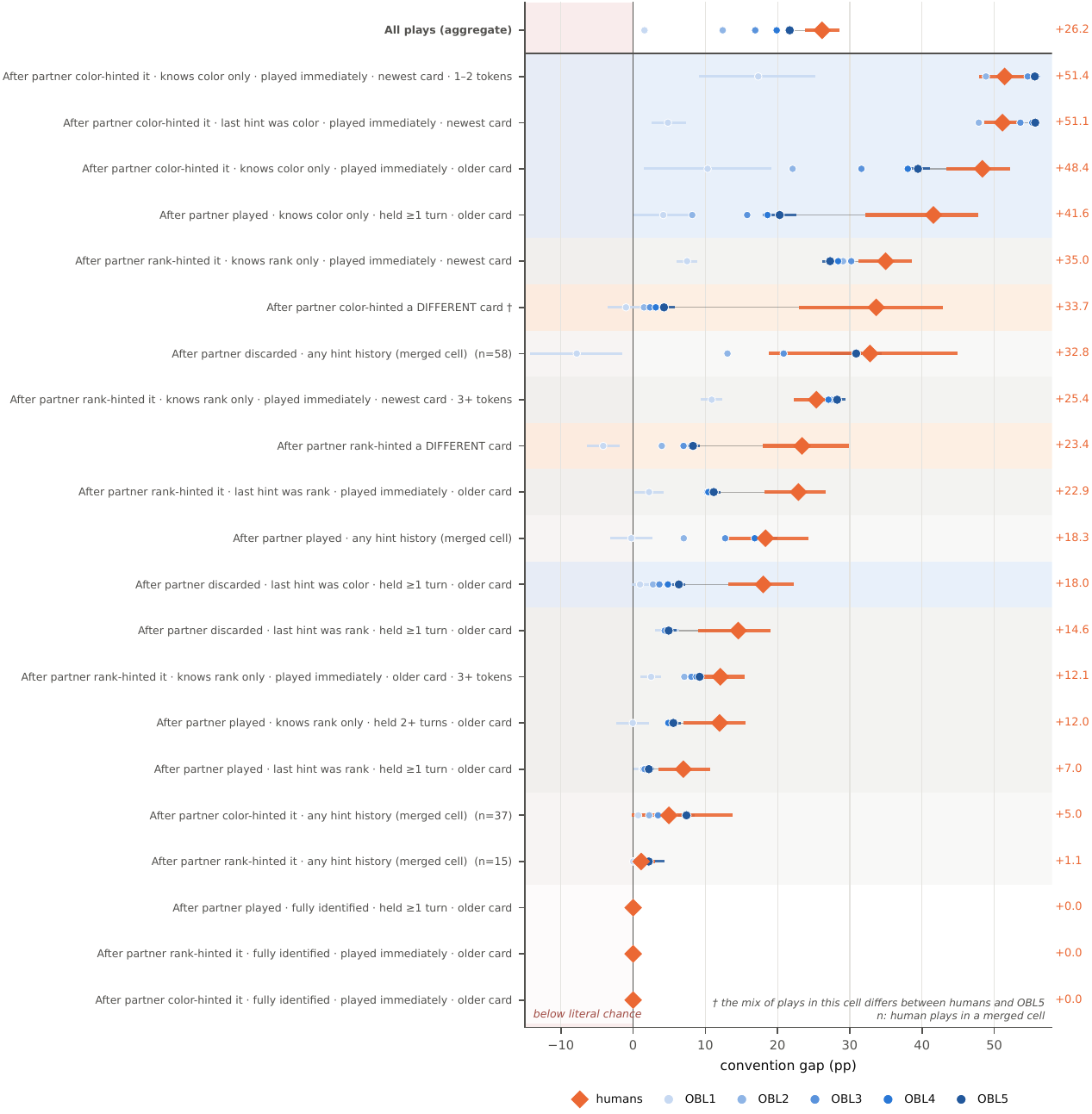}
  \caption{\textbf{Per-category convention gap for all six populations on the shared
  taxonomy.} Rows are the cells of the shared taxonomy (\S\ref{sec:where}); a row exists only if every population has enough plays in it. Markers are estimates on the estimation halves; bars are 95\% bootstrap intervals (human pairs or agent games resampled). The right column gives each row's human value. The top row is the corpus-level aggregate over all plays
  ($\hhGapJointTwoTier\pp$ humans; $\gapOBLOne$ to $\gapOBLFive\pp$ for levels 1--5).
  $\dagger$ marks the one cell where the mix of plays differs between humans and level~5 (\suppfig~\ref{fig:composition}); $n$ is the number of human plays in a merged cell. Row shading: indirect reference (orange), color known (blue), rank known (gray), merged (light gray), fully identified (unshaded); the strip left of zero marks gaps below literal chance.
  \emph{Reading the row labels:} each row is one play context, described left to right
  by (i)~the partner's action just before the play (``After partner color-hinted it'':
  a color hint that touched the played card; ``\ldots color-hinted a DIFFERENT card'': the
  hint touched another card, i.e., indirect reference); (ii)~what hints have revealed
  about the played card (``knows color only'', ``fully identified''; ``last hint was
  color'' when only the most recent hint type is kept); (iii)~timing (``played
  immediately'' after that hint, or ``held ${\geq}1$ turn''); (iv)~hand position (newest
  or older card); and, where it splits the data, (v)~the hint tokens available.
  ``Merged cell'' rows pool rarer contexts that share the listed parts.}
  \label{fig:shared}
\end{figure*}

\begin{figure}[t]
  \centering
  \includegraphics[width=\columnwidth]{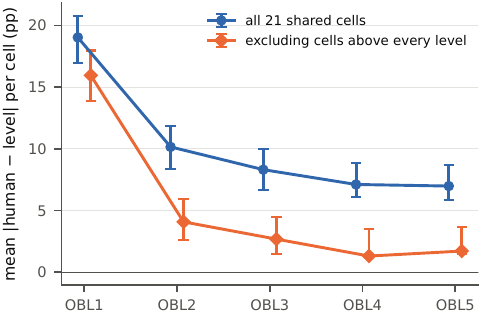}
  \caption{\textbf{Mean per-cell distance between each level's convention-gap
  profile and the human profile.} For each level~$k$, the distance $D_k$: the mean of $|$human $-$ level$|$ over the $\ldNSharedCells$ category rows of Fig.~\ref{fig:shared} (pp), with 95\%
  joint-bootstrap intervals. Blue: all $\ldNSharedCells$ cells. Orange: the
  $\ldNExcludedChannelCells$ cells where the human value exceeds every level's
  (identified by the nearest-level bootstrap of Appendix~\ref{app:depth}) are excluded, leaving the conventions both populations
  share. Intervals are bootstrap percentiles around the full-sample value and are
  asymmetric by construction: resampling noise can only inflate a near-zero per-cell
  difference. The contrasts quoted in the text cancel this bias within draws. This
  per-cell distance is not the difference of the aggregates in the top row of
  Fig.~\ref{fig:shared}.}
  \label{fig:leveldist}
\end{figure}

\subsubsection{Humans sit above the OBL hierarchy, by a small, structured margin}

How far above the OBL hierarchy does human play sit? The human corpus has a gap of
$\hhGapJointTwoTier\pp$, against level~5's $\gapOBLFive\pp$ (Fig.~\ref{fig:shared},
top row; mean $\pm$ SD over the five seed models: $\tSevenSelfplayMeanOBLFive \pm \tSevenSelfplaySdOBLFive\pp$, Appendix~\ref{app:seeds}). Humans are therefore above every level of the hierarchy (level-by-level
view and bot-exclusion sensitivity: Appendix~\ref{app:corpus}).

\paragraph{The aggregate difference is mostly agreement.}
Is the human margin spread across all situations, or concentrated in a few? Measured
cell by cell on the shared taxonomy (\S\ref{sec:where}; Fig.~\ref{fig:shared}), it is
concentrated. On the cells both populations share, the mean per-cell distance $D_k$ between the human profile and the profile of level~$k$ falls from $D_1 = \ldDCoreOnePt$ to $D_2 = \ldDCoreTwoPt\pp$. On these cells, humans are not significantly closer to
level~5 than to level~2 ($D_2 - D_5 = \ldDCoreTwoMinusFive\pp$; Fig.~\ref{fig:leveldist}).
(Including the $\ldNExcludedChannelCells$ cells where the human value exceeds every level's, the distance keeps falling slightly; Appendix~\ref{app:depth}.) What remains of the human
margin sits in a few rare cells: ranking the cells by the human--level-5 difference
puts the two indirect-reference cells and a delayed color cell at the top
(\suppfig~\ref{fig:shareddiff}).

A Kitagawa--Oaxaca--Blinder split \citep{kitagawa1955,oaxaca1973,blinder1973} of the shared-cell difference ($\sharedTaxAggHuman$ against $\sharedTaxAggOblFive\pp$, aggregates over the $\sharedTaxCells$ shared cells on the estimation halves) assigns
$\sharedTaxRate\pp$ to gap differences \emph{within} cells, weighted by human shares,
and $\sharedTaxMix\pp$ to differences in cell shares, weighted by level-5 gaps. The
split is an accounting identity. With the ranking above, the two populations are \emph{similar on
everything frequent and differ in kind in a small family of rare situations}.

\paragraph{The gap is not skill.}
Could the human margin simply reflect stronger play? Level~5 \emph{outscores} the average human game ($\scoreOBLFive$ against the corpus
mean of $\hhMeanScore$), while humans have the larger gap. The gap says how far plays depart
from their literal prediction, not how high a player scores. Two score-selected corpora point the same way. The AH2AC2 corpus, whose selection rule puts its mean score ($\axMeanScore$) close to level~5's, has a gap of $\axGapJoint\pp$, and hanab.live games above the same score floor give $\hhScoreFloorGap\pp$ (Appendix~\ref{app:corpus}). Both select on final score, and so on failed plays, so they describe these games rather than control for skill. Because the platform's community publishes and teaches a convention system, the estimate describes taught play, not typical human play.

\subsubsection{The human profile replicates on an independently collected corpus}
\label{sec:replication}

We next replicated the profile on a second corpus. The AH2AC2 benchmark \citep{ah2ac2} independently released
$\axNGamesReleased$ two-player games from the same platform and community; the two
collections share no game records. Replayed through the identical pipeline, this
corpus has a gap of $\axGapJoint\pp$ against the hanab.live corpus's $\axOursGapJointGameCl\pp$, both
clustered by game (AH2AC2 has no player identifiers, so pair clustering is not
possible). Its category profile correlates with that of the hanab.live corpus at
$r = \axPearsonVsOursExclZero$, and it shows the same indirect-reference excess. Its differences are consistent with its selection rule, and on the sub-population the two corpora share, the cell values agree closely ($\axColorKnownSubcellGapPt$ against $\tTwelveAColorOnlyGap\pp$ on color-known cards). The structure and the indirect-reference excess are properties of the community's play, not of how the hanab.live corpus was built (Appendix~\ref{app:corpus}).

\subsubsection{Where no level approaches the human gap}

On the situations both populations share, humans match level~5 where it plays most often (fresh color-hint plays,
$\hhSitFreshColorGap\pp$ against $\hhSitOblFiveFreshColorGap\pp$) and exceed it
modestly elsewhere. One situation is different in kind.

The two populations differ not only in a situation's gap but in how often they enter it at all
(Table~\ref{tab:usage}). The dominant shared situation is a mirror image: level~5
puts $\tFiveShareFreshColorOBLFive\%$ of its plays into the fresh color-hint
response, against $\hhSitFreshColorShare\%$ for humans. Indirect reference shows the opposite pattern: humans enter it $\hhCatIndirectColorShare\%$ of the time against level~5's $\oblFiveCatIndirectColorShare\%$, and they play unhinted cards $\hhSitUnhintedNewestShare\%$ (newest card) and $\hhSitUnhintedOtherShare\%$ (older cards) of the time, where level~5 almost never does. Level~5 does not merely have a smaller gap in these situations;
it seldom enters them.

\begin{table*}[t]
  \centering\footnotesize
  \caption{\textbf{How often each convention situation is entered} (share of plays, \%),
  with the situation's convention gap (pp) in parentheses. Human columns: the hanab.live corpus
  (all skill levels) and the independently collected AH2AC2 corpus
  (score-selected). Indirect-reference rows use the frozen level-5 categories of \S\ref{sec:where}; for OBL1 they give shares only. Gaps are estimates; intervals for the values discussed are in the text.}
  \label{tab:usage}
  \begin{tabular}{lcccc}
  \toprule
  Situation & OBL1 & OBL5 & Humans (hanab.live) & Humans (AH2AC2) \\
  \midrule
  Fresh color hint & $\tFiveShareFreshColorOBLOne$ ($\tFiveGapFreshColorOBLOneTrend$) & $\tFiveShareFreshColorOBLFive$ ($\tFiveGapFreshColorOBLFiveTrend$) & $\hhSitFreshColorShare$ ($\hhSitFreshColorGapPt$) & $\axSitFreshColorShare$ ($\ptof{\axSitFreshColorGap}$) \\
  Fresh rank hint & $\tFiveShareFreshRankOBLOne$ ($\tFiveGapFreshRankOBLOnePt$) & $\tFiveShareFreshRankOBLFive$ ($\tFiveGapFreshRankOBLFivePt$) & $\hhSitFreshRankShare$ ($\hhSitFreshRankGapPt$) & $\axSitFreshRankShare$ ($\ptof{\axSitFreshRankGap}$) \\
  Hinted earlier & $\tFiveShareStaleHintedOBLOne$ ($\tFiveGapStaleHintedOBLOnePt$) & $\tFiveShareStaleHintedOBLFive$ ($\tFiveGapStaleHintedOBLFivePt$) & $\hhSitHintedEarlierShare$ ($\hhSitHintedEarlierGapPt$) & $\axSitStaleHintedShare$ ($\ptof{\axSitStaleHintedGap}$) \\
  Unhinted, newest & $\tFiveShareUnhintedNewestOBLOne$ ($\tFiveGapUnhintedNewestOBLOnePt$) & $\tFiveShareUnhintedNewestOBLFive$ ($\tFiveGapUnhintedNewestOBLFivePt$) & $\hhSitUnhintedNewestShare$ ($\hhSitUnhintedNewestGapPt$) & $\axSitUnhintedNewestShare$ ($\ptof{\axSitUnhintedNewestGap}$) \\
  Unhinted, older & $\tFiveShareUnhintedOtherOBLOne$ ($\tFiveGapUnhintedOtherOBLOnePt$) & $\tFiveShareUnhintedOtherOBLFive$ ($\tFiveGapUnhintedOtherOBLFivePt$) & $\hhSitUnhintedOtherShare$ ($\hhSitUnhintedOlderGapPt$) & $\axSitUnhintedOtherShare$ ($\ptof{\axSitUnhintedOtherGap}$) \\
  \midrule
  Indirect ref.\ (color) & $\oblOneCatIndirectColorShare$ & $\oblFiveCatIndirectColorShare$ ($\hhCatIndirectColorOblFive$) & $\hhCatIndirectColorShare$ ($\hhCatIndirectColorHumanPt$) & $\axCatIndirectColorShare$ ($\axChanIndirectColorPt$) \\
  Indirect ref.\ (rank) & $\oblOneCatIndirectRankShare$ & $\oblFiveCatIndirectRankShare$ ($\hhCatIndirectRankOblFive$) & $\hhCatIndirectRankShare$ ($\hhCatIndirectRankHumanPt$) & $\axCatIndirectRankShare$ ($\axChanIndirectRankPt$) \\
  \bottomrule
  \end{tabular}
\end{table*}

\paragraph{Indirect reference: a taught convention, rare in level~5's play.}
Where the partner's hint touched a \emph{different} card, humans reach
$\hhCatIndirectColorHuman\pp$ against level~5's $\hhCatIndirectColorOblFive\pp$
($n = \hhCatIndirectColorN$ human plays); for rank hints the values are
$\hhCatIndirectRankHuman$ against $\hhCatIndirectRankOblFive\pp$. The sharpest case
is an unhinted card played immediately after a color hint that touched another card
($n = \silHHAllColorOtherN$). There, the literal failure probability is $\silHHAllColorOtherPost\%$, but the actual failure rate is only $\silHHAllColorOtherLoss\%$ (\supptab~\ref{tab:silencesplit}). Level~5 rarely makes this play, and its gap there is small (\suppfig~\ref{fig:composition}). The human pattern matches a taught convention (the
community's \emph{finesse}) in which a hint about one card tells the partner to play
a different card from an agreed slot (\S\ref{sec:discussion} discusses why the
hierarchy may lack it).

\paragraph{Unhinted cards.} Humans reach $\hhSitUnhintedNewestGap\pp$ on unhinted newest cards and $\hhSitUnhintedOlderGap\pp$ on older ones, against $\hhSitOblFiveUnhintedNewestGap$ and $\tFiveGapUnhintedOtherOBLFivePt\pp$ for level~5. Most of these human plays ($\silHHAllHintOtherN$ of $\silHHAllAllN$) follow a hint to another card, which places them in indirect reference. On the remaining plays, mostly after a partner play or discard, the human gap is $\silHHAllCleanGap\pp$ against level~5's $\silOBLFiveAllCleanGapPt\pp$, a difference of $\silDiffHHMinusOBLFiveAllClean\pp$ (\supptab~\ref{tab:silencesplit}).

\paragraph{The largest differences survive matching on card configuration.}
Indirect reference sits in merged cells of the shared taxonomy, which collect the finer
situations that did not reach every population's $n_{\min}$. A single label can
therefore hold different plays for different populations, so we asked whether the
split survives a finer classification. Splitting the two indirect-reference cells by
the played card's own hint history shows that it does
(\suppfig~\ref{fig:composition}), and the difference has three parts that all point
the same way. \emph{Occupancy}: humans enter ``after the partner
color-hinted a different card'' mostly holding \emph{unhinted} cards, the finesse
configuration, while the hierarchy's rare entries hold rank-known or fully-identified
cards. The human-specific sub-cells are nearly empty for every agent level, and that
emptiness is itself a finding. \emph{Rate on matched sub-populations}: where both
populations do occupy a sub-cell, the difference remains. On the finesse sub-cells
humans read $\tTwelveANeverHintedGap$ and $\fcRnkHumanUnhintedGapPt\pp$ against
level~5's $\fcColOblFiveUnhintedGapPt$ and $\fcRnkOblFiveUnhintedGapPt\pp$; on the color-known sub-cell, $\tTwelveAColorOnlyGap$ against
$\fcColOblFiveColorOnlyGapPt\pp$ with non-overlapping intervals; and this color-known sub-cell agrees closely across the two corpora (\S\ref{sec:replication}). \emph{Configuration}: which cards a population
holds when it acts in these situations suggests a different rule for when such a
play is allowed. (The merged shared-taxonomy cells of Fig.~\ref{fig:shared} average
over this composition, which is why their values sit below the frozen-cell values
quoted here.) Thus the finer classification does not dissolve the finding; it
sharpens it.

%% file: sections_aamas/04_when.tex
\subsection{Where the conventions are used}
\label{sec:when}

To locate the gap, we attach to each play its context (turn, the hint history
of the played card, the partner's immediately preceding action, hand position, and
token state) and ask when, in which situations, and in which feature of the game state it
sits.

\begin{figure*}[t]
  \centering
  \includegraphics[width=\textwidth,height=0.24\textheight,keepaspectratio]{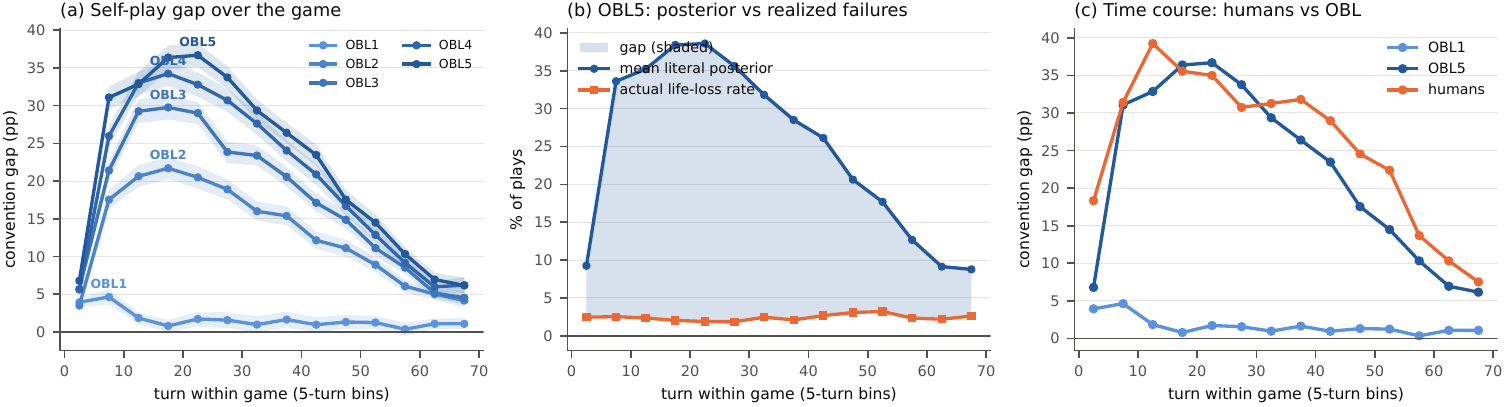}
  \caption{\textbf{Convention gap by turn within the game.} (a)~Convention gap by
  5-turn bin for levels 1--5 in self-play (both players' plays; bins
  with $\geq 500$ plays; bands are 95\% game-clustered bootstrap intervals). The
  inverted U rises with level, peaking at $\tcfigPeakGapOBLTwo\pp$ (level~2) to
  $\tcfigPeakGapOBLFive\pp$ (level~5) in the mid-game. (b)~Level~5 decomposed: the mean
  literal failure probability of the chosen plays against the realized life-loss rate;
  the shaded area is the gap. The realized rate is flat; the literal price of the plays
  is what moves. (c)~Humans against levels~1 and~5 by 5-turn bin (bins with $\geq 200$ human plays); lines are estimates without intervals.}
  \label{fig:timecourse}
\end{figure*}

\subsubsection{The gap peaks in the mid-game}

We first broke the gap down by turn. For levels 2--5 the gap follows an inverted~U over the game: close to zero in the
opening ($\tcOpeningGapOBLTwo$ to $\tcOpeningGapOBLFive\pp$ in turns 0--4), peaking in
the mid-game ($\tcPeakGapOBLFive\pp$ at level~5, in the 5-turn bin starting at turn
$\tcPeakTurnBinOBLFive$), then falling to between $\tcEndgameGapOBLTwo$ and
$\tcEndgameGapOBLFive\pp$ late (Fig.~\ref{fig:timecourse}a). Level~1 stays low
throughout (\dotrange{\tcOBLOneRangeAllBins}\pp{} across all bins). At level~5 the shape comes
from the literal risk of the chosen plays: the actual failure rate is flat at
\hyrange{\tcfigOblFiveLossRateRange}\% across all bins, while the mean literal failure
probability rises and falls
(Fig.~\ref{fig:timecourse}b). In this sense conventions are used most in the
mid-game: there the plays carry the most literal risk yet fail no more often.

Human play follows the same inverted U (Fig.~\ref{fig:timecourse}c): similar at the
peak ($\tcfigHumanPeakGap$ against level~5's $\tcPeakGapOBLFive\pp$), but above
level~5 in the opening bin ($\tcfigHumanOpeningGap$ against
$\tcOpeningGapOBLFive\pp$) and in every late bin from turn~40 on
(\slashof{\tcfigHumanLateBinsAboveOblFive} bins). The human gap rises earlier and stays high
longer than the agents'.

\subsubsection{One situation contributes most of level~5's gap}

We next sorted every play into the five hand-crafted situations. One situation
dominates: a card played immediately after the partner color-hinted it. A color
hint gives the suit and says nothing about the rank, so playing at once is reckless
by the literal reading. At level~5 such plays have a mean literal failure probability
of $\tFivePostFreshColorOBLFive\%$ but fail only $\tFiveLossFreshColorOBLFive\%$ of
the time, a gap of $\tFiveGapFreshColorOBLFive\pp$. The situation grows from
$\tFiveShareFreshColorOBLOne\%$ of plays at level~1 to
$\tFiveShareFreshColorOBLFive\%$ at level~5, while its gap grows from
$\tFiveGapFreshColorOBLOneTrend$ to $\tFiveGapFreshColorOBLFiveTrend\pp$
(Table~\ref{tab:usage}). That one situation contributes $\tcFreshColorShareOfGapOBLFive\%$ of level~5's total gap (share $\times$ gap in Table~\ref{tab:usage}, over the level-5 total). To a first approximation,
the rise of the gap with level reflects both partners' step-by-step adoption of one
rule: a color hint marks a card as playable.

Unhinted plays change sign across the OBL hierarchy. They are rare and, read literally, risky. Level~1 loses more often than literal chance predicts
($\tFiveGapUnhintedNewestOBLOne\pp$ on the newest card), while level~5 reaches $\tFiveGapUnhintedNewestOBLFive\pp$ in the same
situation. One reading is that the absence of a warning carries information: had the newest card been dangerous, the partner might have acted. In human play, most unhinted plays instead follow a hint to another card (\S\ref{sec:what}).

\subsubsection{At level~5 the gap tracks what the played card is known to be}

We then read the induced categories across levels to see which feature of the game
state carries the gap (full cross-level reading: Appendix~\ref{app:metric}). At level~5 the induced categories explain $R^2 = \tNineRsqBackoffOBLFive$ of the per-play residual variance out of sample, against $\tNineRsqFiveOBLFive$ for the five hand-crafted situations (\suppfig~\ref{fig:varexpl}). The
carrier moves. At levels 2--3 the main split is hand position; by level~4 it is the
played card's hint history; at level~5 it is color-only knowledge itself, \emph{what
the card is known to be} rather than how fresh the hint was. The two highest-gap
categories in the whole taxonomy are \emph{delayed} color-only plays, played two to
three actions after the hint ($\tOneZeroGapOne\pp$) and four or more
($\tOneZeroGapTwo\pp$), above the largest fresh-hint categories (\suppfig~\ref{fig:taxonomy}).
The hand-crafted map filed these plays under \emph{hinted earlier} at a diluted
$\txStaleHintedOBLFive\pp$. Thus the hint, or the later absence of a warning, carried
information beyond literal content.

%% file: sections_aamas/06_worth.tex
\subsection{What removing conventions costs}
\label{sec:worth}

If high-gap channels mark conventions the policy relies on, then blocking them
should cost more score than blocking equally frequent plays at random
(design in \S\ref{sec:maskdesign}). It does, but only when several channels are blocked together, and \emph{how}
it does is the finding: the taxonomy ranks categories by their
contribution to the gap, and that ranking does not predict what removing a
channel costs.

\begin{figure*}[t]
  \centering
  \includegraphics[width=\textwidth]{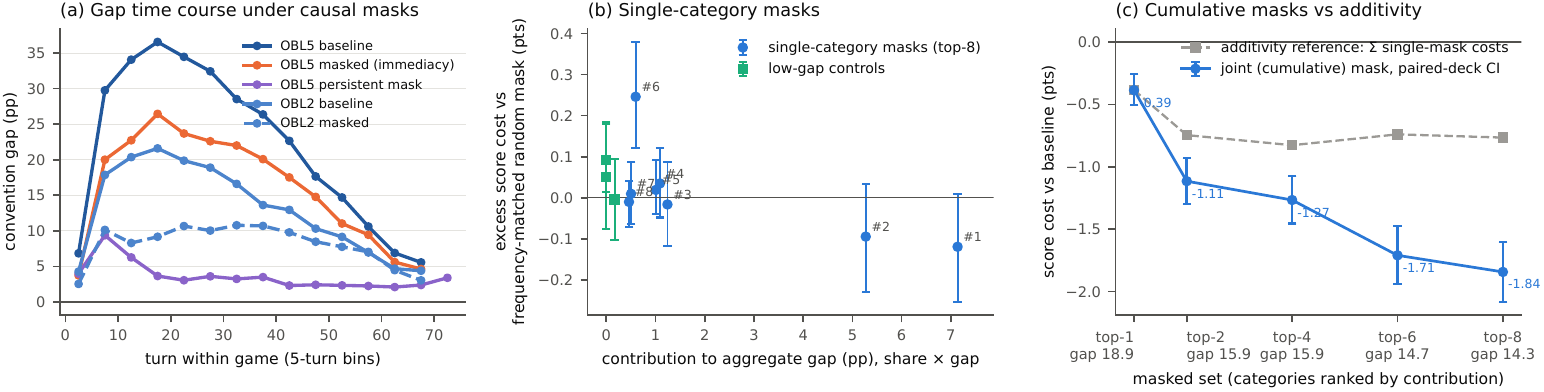}
  \caption{\textbf{Gap and score effects of the masking interventions.}
  (a)~Gap time course under the whole-channel masks: level~5 baseline, immediacy mask
  and persistent mask; level~2 baseline and mask. The immediacy mask scales the
  mid-game peak down while preserving its shape
  ($\tcfigMaskPeakOblFiveBaseline \rightarrow \tcfigMaskPeakOblFiveImmediacy\pp$; the
  blocked plays reappear later), whereas the persistent mask removes the peak
  ($\tcfigMaskPeakOblFivePersistent\pp$).
  (b)~Excess score cost of each single-category mask (the paired score cost of
  masking the category minus the cost of a frequency-matched random mask at the same
  trigger rate) against the category's contribution to the aggregate gap (its share
  of plays times its gap, in pp), with low-gap controls in a second color.
  (c)~\emph{Raw} paired score cost of masking the top-$k$ categories jointly, against
  the additivity reference (the sum of the $k$ single-mask costs); raw costs rather
  than excess, so the super-additivity does not depend on the null-curve fit; the second line of each x-axis label gives the total gap under that mask (pp; baseline $\eFiveBaselineGap$). In (b) and (c), bars are 95\% paired-deck bootstrap intervals. One category costs little; several together cost more than the sum of their single costs.}
  \label{fig:intervention}
\end{figure*}

\subsubsection{Blocking the color-hint response reroutes the convention rather than destroying it}

We first masked the OBL hierarchy's dominant convention as a whole. Blocking the
immediate response to a partner color hint, for both players, changes score by
$\tEightOblFiveMaskBothDeltaScore$ points at level~5
($\tEightOblFiveBaseScore \rightarrow \tEightOblFiveMaskBothScore$). Blocking it for
one player has a little under half that effect ($\tEightOblFiveMaskPZeroDeltaScore$), and only
the blocked player's gap moves ($\tEightPZeroMaskedPlayerGapBase \rightarrow
\tEightPZeroMaskedPlayerGapMask\pp$, against
$\tEightPZeroUnmaskedPlayerGapBase \rightarrow \tEightPZeroUnmaskedPlayerGapMask\pp$
for the partner; Appendix~\ref{app:interventions}).

But the blocked convention is not destroyed; it is rerouted. Under the immediacy mask
the displaced plays reappear one turn later: the \emph{hinted earlier} share doubles
($\tEightStaleShareBase\% \rightarrow \tEightStaleShareMask\%$) and its gap triples
($\tEightStaleGapBase \rightarrow \tEightStaleGapMask\pp$) at an unchanged loss
rate. The convention's content survives; only its timing breaks, and the score loss
is consistent with a cost of delay. Removing these plays instead, by forbidding the play of
color-hinted-only cards altogether, changes score by $\tEightOblFivePersistDeltaScore$ points
and collapses the gap to $\tEightOblFivePersistGap\pp$, with the replaced actions
shifting toward discards ($\tEightPersistDiscardShare\%$; Appendix~\ref{app:interventions}). The two masks leave
different marks on the time course (Fig.~\ref{fig:intervention}a). The same immediacy mask at level~2 changes
score by $\tEightOblTwoMaskBothDeltaScore$: the cue matters there too, but less.

\subsubsection{Single masks cost little beyond generic disruption}

We next asked whether the observational ranking predicts the cost of removal. We
masked each of the top eight categories by observational contribution (share
$\times$ gap) one at a time (Fig.~\ref{fig:intervention}b). The three low-gap control categories give excess costs of $\tOneFourExcessCtrlA$, $\tOneFourExcessCtrlB$ and $\tOneFourExcessCtrlRlagFour$ points, so the control curve is accurate to about a tenth of a point: the size of the single-mask excesses below, and an order of magnitude below the joint-mask interaction of \S\ref{sec:worth}. Only the two
largest-contribution categories have negative excess estimates ($\tOneFourExcessSingleOne$
and $\tOneFourExcessSingleTwo$ points), and both intervals reach zero. Apart from
category~\#6, the others lie between $\tOneFourExcessSingleThreePt$ and
$\tOneFourExcessSingleFourPt$. With eight categories, two of which carry most of the
leverage, we do not read a dose--response relation into them (Appendix~\ref{app:interventions}).
Even the top category is mostly generic disruption. Category~\#1 holds
$\eFiveTopCellSharePct\%$ of plays at $\eFiveTopCellGap\pp$, and masking it changes
score by $\tOneFourCostSingleOne$ points raw, but a random mask at the same trigger
rate already changes it by $\eFiveTopCellRandomCost$ (\supptab~\ref{tab:conditions}); the excess specific to the
convention is only $\tOneFourExcessSingleOne$. Thus most of a single mask's cost is
disruption, not lost meaning.

Masking category~\#6 raises the score beyond the random mask (excess $\tOneFourExcessSingleSix$ points). A pre-registered out-of-sample replication gives a smaller excess ($\eFivebPooledExcess$ pooled; Appendix~\ref{app:interventions}). We read this as no detectable positive value for this convention, with the first estimate inflated by selecting among eight categories.

\subsubsection{Joint masks cost far more than the sum of single masks}

If single masks are cheap because meaning reroutes, then closing the substitutes
together should cost more than their sum. It does. Masking the top eight categories
together changes score by $\eFiveCumTopEightCost$ points,
against $\eFiveSumSinglesTopEight$ for the sum of the eight single masks, a super-additive interaction of $\eFiveSuperAddTopEight$ points (joint minus summed; Fig.~\ref{fig:intervention}c), and lowers the total gap by a third, from $\eFiveBaselineGap$ to $\tOneFourTotalGapCumEight\pp$ (\supptab~\ref{tab:conditions}). This is consistent with rerouting: each single mask
leaves substitute channels open, and joint masks close them together (the
redistribution is in Appendix~\ref{app:interventions}).

The interventions push both agents off their training distribution, so these costs
are upper bounds under retraining: they price a channel for this fixed pair of
policies.

%% file: sections_aamas/07_discussion.tex
\section{Discussion}
\label{sec:discussion}

In this study, we placed human play and the OBL hierarchy on one
scale, broke the convention gap down by situation and by turn, and intervened on the
agents' policies against a frequency-matched control. We found that the two
populations behave alike in frequent situations, differ in indirect reference, where no level approaches the human gap, and that a convention's gap contribution does not predict what its
removal costs.

\paragraph{Are humans and the OBL hierarchy different?}
On the conventions the populations share (everything frequent), they are similar cell
by cell, and humans are not significantly closer to level~5 than to level~2. The differences that survive every control are differences
in \emph{kind}, not only in size. Indirect reference shows a human gap that no level approaches: $\tTwelveAColorOnlyGap$ against level~5's $\fcColOblFiveColorOnlyGapPt\pp$ on the color-known sub-cell, replicated across two independently collected corpora. On unhinted cards that do not follow a hint to another card, the human gap differs from level~5's by $\silDiffHHMinusOBLFiveAllClean\pp$. Humans also enter indirect reference far more often, and their gap rises earlier and stays high longer. The score-selected AH2AC2 corpus shows the same pattern, and clones trained on human play keep the shared conventions while losing the human-specific cells (Appendix~\ref{app:clones}). The populations have similar gaps on the conventions they share and differ on indirect reference.

\paragraph{Gap contribution is not importance.}
The split in \S\ref{sec:worth} may apply beyond Hanabi. Removing the eight
highest-contribution conventions cuts the total gap by a third but costs
under two points of score, consistent with rerouting. The cost
appears only when the substitutes are closed together, and then it is more than
additive. A channel's gap contribution and how hard it is to replace are
therefore different quantities. Any interpretability method that ranks channels by
an observational measure and reads that ranking as importance may inherit
this problem; the fix is joint ablation against a matched null (\S\ref{sec:worth}).

\paragraph{What ZSC gets, and what it does not.}
Off-belief learning's level~1 is trained to read past actions only for their literal
content, which needs no shared convention. Human players appear to coordinate the opposite way: they arrive already sharing a taught system (Appendix~\ref{app:community}), and their gap rises earlier in the game than the agents' (\S\ref{sec:when}). Indirect reference suggests what is missing: a representation of hints as speech acts about the whole hand \citep{frank2012rsa} rather than as signals about the touched card. Off-belief learning reads each past
action as if a fixed base policy had produced it, so a hint on card~A carries only
what that policy put into it, and level~5 seldom plays card~B after it. Depth alone does not account for the difference (Appendix~\ref{sec:depth}). One interpretation is that a hierarchy whose beliefs rest on a fixed base policy cannot express such a hint.

\paragraph{Supporting analyses.}
\Suppmat{} adds four results: the realized gap depends on the partner (lower level
in seat~0; Appendix~\ref{app:mixed}); the agent-grown taxonomy transfers to human play
(Appendix~\ref{app:corpus}); depth of reasoning does not separate the populations
(Appendix~\ref{app:depth}); and clones of both populations reproduce the split between shared conventions and human-specific cells (Appendix~\ref{app:clones}).

\paragraph{What the estimates cover.}
The category analysis is descriptive: categories are defined by the action the policy
chose, so a category's gap says how far its plays depart from the literal prediction, not what would
happen if the situation were created on purpose. \S\ref{sec:worth} supplies the
causal complement for the agents' own conventions; indirect reference cannot be priced this way, because no level enters it often enough to mask. The human corpus comes from a community that publishes and teaches a convention system, so its estimates describe taught play, not typical human play; the pattern replicates on a second, score-selected corpus. All of this is
one game; the method needs only an exact literal baseline and a maskable policy.

\subsubsection*{Reproducibility}
The metric implementation is public, released with the previous study \citep{paper1}, and used at a pinned
commit. Every game is replay-verified against its ground-truth score, all numbers were
recomputed by a separate audit pipeline, and all our intervals but one bootstrap
(Appendix~\ref{app:repro}) reproduce exactly from recorded seeds.

%% file: appendix/A_metric.tex
\section{The metric and the category-induction method}
\label{app:metric}

This appendix defines the literal observer behind \maintext Eq.~\ref{eq:gap}, the context
descriptor and cell-size rule used to induce categories, and how the literal posterior
rises with level.

\subsection{The literal observer and a worked example}
\label{app:metric:worked}

At a play action, the literal observer knows, for every card in the acting player's hand,
the set of identities consistent with all positive and negative hint constraints received
so far, and it counts copies over every visible zone (the partner's hand, the discard
pile, the fireworks). Because a hint that touches one card constrains the identities
available to another, the observer computes the \emph{joint} distribution over the whole
hand, weighting each joint assignment by the number of ways to draw it without
replacement from the hidden pool, and marginalizes to the played card. The probability
that the played card is unplayable under this distribution is the play's literal failure
probability $P(\text{life loss})$; Eq.~\ref{eq:gap} averages it over a set of plays and
subtracts the realized failure rate.

A schematic example with five plays illustrates the arithmetic (the values here are
illustrative, not data). Suppose the literal failure probabilities of five played cards
are $0.40$, $0.40$, $0.40$, $0.20$ and $0.60$, and exactly one of the five plays fails.
The mean literal failure probability, the \emph{literal price} of the plays, is $0.40$; the realized failure rate is $0.20$; the gap is
$+20$ percentage points: the plays succeeded more often than their literal information
justified. A gap near zero does not mean the players were well informed, only that
their success rate matches the literal price of the plays they chose. The gap is signed:
plays can beat or miss their literal price. It cannot exceed the mean literal price.

Two properties matter for everything downstream. First, the baseline is exact:
there is no learned component, so a decomposition of the aggregate (by turn, by
situation, by category) is free of model error. Second, the
observer conditions on \emph{observation content but not on action choice}: the fact that
the partner chose this hint at this moment carries no weight. What the choice
itself conveys can show up in the gap, but need not: subsets of plays with gaps of opposite
sign can cancel, and plays whose outcome literal information fixes give a gap of exactly zero.

\subsection{The six-component descriptor}
\label{app:metric:descriptor}

Every play is described by six components of its context:
\begin{enumerate}\itemsep1pt
  \item \textbf{Partner's immediately preceding action}: play, discard, color hint or
        rank hint, each hint distinguished by whether it touched the played card
        (\emph{touch}) or a different card (\emph{other}), or no preceding action (first
        move).
  \item \textbf{Hint relation to the board} (hints only): whether the hinted color's
        stack is open, or the hinted rank is currently needed on some stack.
  \item \textbf{Hint history of the played card}: the ordered pattern of hint types that
        ever touched it (none, C, R, CR, RC, CC+, RR+, or mixed, meaning three or more).
  \item \textbf{Lag}: actions since the last hint that touched the card ($1$, $2$--$3$,
        $4{+}$, or unhinted). In two-player play only odd lags occur, so lag $2$--$3$
        means exactly one own action between the hint and the play.
  \item \textbf{Hand position}: newest card or older.
  \item \textbf{Hint-token stock}: $0$, $1$--$2$, or $3{+}$ tokens.
\end{enumerate}
The cross-product yields \txTypesRange{} observed context types per level with a
heavy-tailed frequency distribution; the number of types needed to cover $90\%$ of
plays is, per level, \txKNineZero.

\subsection{Cell size, backoff, and the split}
\label{app:metric:nmin}

The minimum cell size is derived, not chosen. With per-play residual standard deviation
$s$ (empirically \txSdRange{} across levels, discovery half), estimating a cell mean to
$\pm 5\pp$ at $95\%$ confidence requires $n_{\min} = \lceil (1.96\,s/0.05)^2 \rceil$,
giving \txNminRange{} from level~1 to level~5. A $\pm 3\pp$ target would demand cells
of $\nMinThreeTargetOBLFive$ plays at level~5. Types that large cover only
\txCovThreepp{} of plays, below the $90\%$ coverage we required, so more plays would
fall into merged cells.

Types below $n_{\min}$ merge by a fixed backoff ladder: drop tokens, then coarsen
lag, then drop the board relation, then collapse hint history to its last type, then
drop position, leaving at worst the partner's preceding action alone. No play
therefore falls into an opaque \emph{other} bucket, and every merged cell retains a
readable label. Categories
and the CART tree are fit on a discovery half (games split 50/50 within level and seed,
with a recorded split seed); every reported statistic (cell
means, intervals, FDR flags, $R^2$) comes from the held-out estimation half
(Table~\ref{tab:playcounts}). The per-category bootstrap intervals reproduce only up to
bootstrap error (Appendix~\ref{app:repro}).

\begin{table}[H]
  \centering\small
  \caption{Play actions entering the category induction, by level and split half.
  Level~1 has one seed model; levels 2--5 pool five seed models each.}
  \label{tab:playcounts}
  \begin{tabular}{lccc}
  \toprule
  Level & Discovery & Estimation & Total \\
  \midrule
  OBL1 & \tNineaDiscOBLOne{} & \tNineaEstOBLOne{} & \tNineaTotalOBLOne{} \\
  OBL2 & \tNineaDiscOBLTwo{} & \tNineaEstOBLTwo{} & \tNineaTotalOBLTwo{} \\
  OBL3 & \tNineaDiscOBLThree{} & \tNineaEstOBLThree{} & \tNineaTotalOBLThree{} \\
  OBL4 & \tNineaDiscOBLFour{} & \tNineaEstOBLFour{} & \tNineaTotalOBLFour{} \\
  OBL5 & \tNineaDiscOBLFive{} & \tNineaEstOBLFive{} & \tNineaTotalOBLFive{} \\
  \midrule
  All  & \tNineaDiscAll{} & \tNineaEstAll{} & \tNineaTotalAll{} \\
  \bottomrule
  \end{tabular}
\end{table}

\subsection{How the literal posterior rises with level}
\label{app:metric:rise}

The mean literal posterior is a property of which plays a policy selects, not of the
game. In self-play it rises from $\sSixMeanPostOBLOne\%$ (level~1) to
$\sSixMeanPostOBLFive\%$ (level~5); the previous study reports this rise and the same plays
by hint count \citep{paper1}. We split this rise into two simultaneous shifts. \emph{Composition}: higher levels shift plays from
cards touched by two or more hints to cards touched by one. The share of plays on a card touched by exactly one hint
grows from $\sSixShareOneHintOBLOne\%$ to $\sSixShareOneHintOBLFive\%$, while plays
on cards touched by two or more hints shrink from $\sSixShareTwoPlusOBLOne\%$ to
$\sSixShareTwoPlusOBLFive\%$. \emph{Within-stratum}: even among one-hint plays,
higher levels accept literally riskier ones. The mean posterior within that stratum
climbs
from $\sSixPostOneHintOBLOne\%$ to $\sSixPostOneHintOBLFive\%$ while the realized loss
rate of the same plays falls from $\sSixLossOneHintOBLOne\%$ to
$\sSixLossOneHintOBLFive\%$. A Kitagawa split of the rise \citep{kitagawa1955} puts
$\sSixPosteriorRiseCompositionPP\pp$ in composition and $\sSixPosteriorRiseWithinPP\pp$ within
strata, with the composition term evaluated at level-1 stratum posteriors and the
within-stratum term at level-5 shares. This is an accounting identity and assigns no
cause. At these weights the composition term is small, because the shrinking share of
plays on cards no hint touched partially offsets the growing one-hint share.

\subsection{The induced taxonomy: full construction and cross-level results}
\label{app:taxfull}

\subsubsection{Data-induced categories}

The five situations of \S\ref{sec:where} (fresh color hint, fresh rank hint, hinted
earlier, unhinted newest, unhinted older; the \emph{hand-crafted map} below) explain
only $\tNineRsqFiveOBLFive$ of per-play residual variance at level~5. We therefore
induce categories from the data, using the descriptor of \S\ref{app:metric:descriptor} and
the cell-size, backoff and split rules of \S\ref{app:metric:nmin}; all reported
statistics come from the estimation half with game-clustered bootstrap intervals and
Benjamini--Hochberg control at $q = 0.05$ \citep{benjamini1995fdr}.

\begin{figure}[tp]
  \centering
  \includegraphics[width=\columnwidth,trim=192 0 0 620,clip]{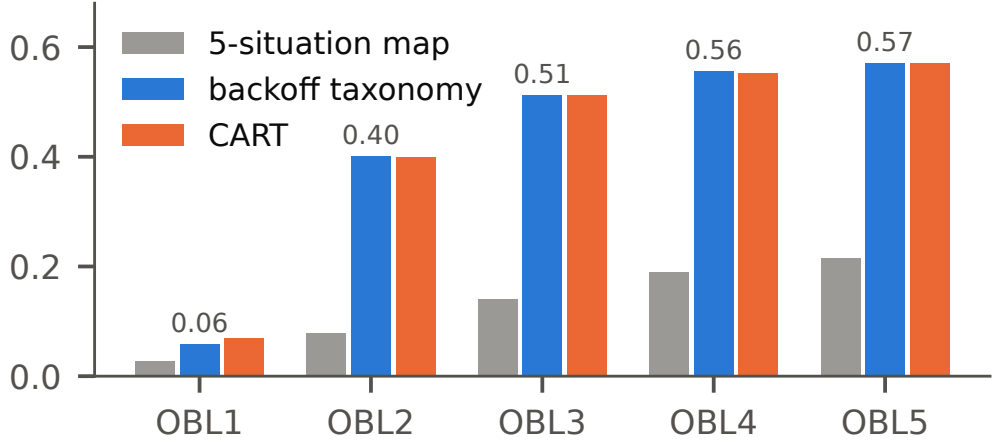}
  \caption{\textbf{Out-of-sample variance explained, per level.} Share of per-play
  residual variance explained on the estimation half (out-of-sample $R^2$, y axis) by the
  hand-crafted five-situation
  map, the backoff taxonomy, and a depth-4 CART tree. All three explain little at level~1
  ($R^2 \le \tNineRsqCartOBLOne$); the backoff taxonomy explains $R^2 = \tNineRsqBackoffOBLFive$
  at level~5.}
  \label{fig:varexpl}
\end{figure}

The induced taxonomy yields $\tNineCatsOBLFive$ categories at level~5 and explains
$R^2 = \tNineRsqBackoffOBLFive$ out of sample, matched by a depth-4 CART tree \citep{breiman1984cart} at
$\tNineRsqCartOBLFive$, against $\tNineRsqFiveOBLFive$ for the hand-crafted map
(Fig.~\ref{fig:varexpl}). At level~1 the taxonomy explains $R^2 = \tNineRsqBackoffOBLOne$
and the CART tree $\tNineRsqCartOBLOne$. The low values do not mean that level~1 lacks
structure. Its tree splits first on whether any hint touched the played card
(Table~\ref{tab:cartroot}), and its plays on cards no hint touched have negative gaps,
$\tFiveGapUnhintedNewestOBLOnePt\pp$ on the newest card and
$\tFiveGapUnhintedOtherOBLOnePt\pp$ on older cards (\maintext Table~\ref{tab:usage}). These plays are
rare ($\tFiveShareUnhintedNewestOBLOne\%$ and $\tFiveShareUnhintedOtherOBLOne\%$ of
level~1's plays), so they can explain little per-play variance.

\begin{figure*}[tp]
  \centering
  \includegraphics[width=\textwidth,height=0.78\textheight,keepaspectratio,
    trim=0 236.4 0 0,clip]{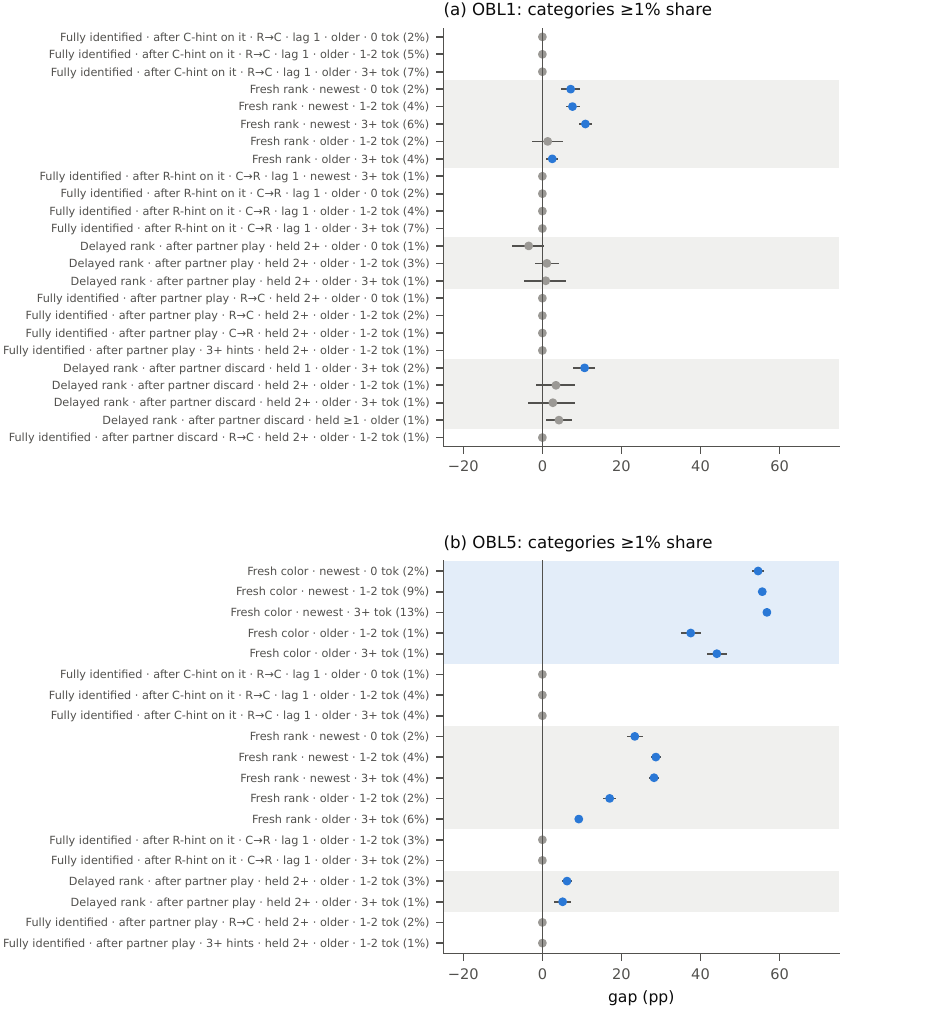}
  \caption{\textbf{Induced-category gap ladders for level~1 (a) and level~5 (b).}
  Categories holding $\geq 1\%$ of plays (share of plays in parentheses). Markers are
  gaps on the estimation half; bars are 95\% game-clustered bootstrap intervals; blue
  markers pass Benjamini--Hochberg control at $q = 0.05$, gray ones do not. Row shading
  gives the information family: color known (blue), rank known (gray), fully identified
  (unshaded). \emph{Reading the row labels,} left to right: the hint type and
  timing (``Fresh color/rank'': played right after a hint of that type touched the card;
  ``Delayed color/rank'': hinted earlier with that type and held before playing; ``Fully
  identified'': color and rank both known, with the order of the hints, e.g.\ R$\to$C for
  rank then color, or ``3+ hints''; ``Finesse-like (color/rank)'', used in later figures:
  played right after a hint of that type touched a \emph{different} card, i.e., indirect
  reference); the partner's action just before the play where it defines the category
  (``after partner play'', ``after partner discard'', ``after C-hint on it''); how long
  the card was held (``held 1'', ``held ${\geq}1$'', ``held 2+'' own turns; ``lag 1'':
  played immediately); hand position (newest or older card); and the hint tokens
  available (``0'', ``1-2'', ``3+ tok'').}
  \label{fig:taxonomy}
\end{figure*}

\begin{table}[tp]
  \centering\small
  \caption{First split of the depth-4 CART tree at each level (discovery half). Hint
  history lists the hint types that touched the played card, in order (C: color, R:
  rank).}
  \label{tab:cartroot}
  \begin{tabular}{ll}
  \toprule
  Level & Root split \\
  \midrule
  1 & Hint history: unhinted vs.\ hinted \\
  2 & Hand position: older vs.\ newest card \\
  3 & Hand position: older vs.\ newest card \\
  4 & Hint history: one color hint only vs.\ any other \\
  5 & Hint history: color hints only vs.\ any other \\
  \bottomrule
  \end{tabular}
\end{table}

\subsubsection{The knowledge state, not the freshness of the hint, carries the gap}

Reading the trees across levels shows the phenomenon relocating (Table~\ref{tab:cartroot}). At levels 2--3 the root
split is hand position; by level~4 it is the played card's hint history; at level~5 it is
color-only knowledge outright. As the OBL hierarchy deepens, \emph{what the card is known to
be} rises from a sub-split to the primary axis of the policy's gap. What carries the gap
is not the freshness of the cue but the knowledge state the cue produced.

The sharpest evidence comes from lag, the number of actions since the last hint that
touched the card. If hint meaning decayed over time, gaps would fall with lag. Instead,
the two highest-gap categories in the
whole level-5 taxonomy are \emph{delayed} color-only plays: the card was color-hinted,
the player did something else first, then played it. The lag-$2$--$3$ category reads
$\tOneZeroGapOne\pp$ and the lag-$4{+}$ category $\tOneZeroGapTwo\pp$, above the two
largest fresh-hint categories at $\tOneZeroGapThree$ and $\tOneZeroGapFour\pp$ (all four
values are level-5 self-play, estimation half). Thus the hint, or the later absence of a warning, carried information beyond literal
content; the gap does not say which. The
hand-crafted map filed all of these under \emph{hinted earlier} at a diluted
$\txStaleHintedOBLFive\pp$, which is how a $\tOneZeroGapOnePt\pp$ category disappears into a $\txStaleHintedOBLFive\pp$ situation.

Two further structures the hand-crafted map misses. Hand position separates otherwise
matched fresh-color categories by \beforeparen{\txNewestVsOlder}{} (levels 2--5), and
token scarcity shifts the same situation by a few points
(\beforeparen{\txTokenModulationOBLFive}{} tokens). Despite roughly a hundred
categories per level, a handful of categories carry $80\%$ of each level's total gap
(per level: \txConcentration). Seed-to-seed differences sit in the tail of the
taxonomy (Appendix~\ref{app:seeds}).

This analysis is descriptive: categories condition on the action the policy chose, so a
category's gap says how far its plays depart from the literal prediction, not what would
happen if the situation were induced. \S\ref{sec:worth} supplies the causal complement.

%% file: appendix/B_mixed.tex
\section{Mixed-level play}
\label{app:mixed}

This appendix gives the analysis behind the statement in \S\ref{sec:discussion} that
the realized gap depends on the partner. \S\ref{sec:who} reports all ten
mixed-level pairings, \S\ref{app:mixed:cue} localizes the effect in the fresh
color-hint cue, and Appendix~\ref{app:seeds} repeats the per-agent comparison of
\S\ref{sec:who} across seed combinations.

\subsection{Per-agent gaps in mixed-level play}
\label{sec:who}

\begin{figure*}[tp]
  \centering
  \includegraphics[width=\textwidth,height=0.8\textheight,keepaspectratio]{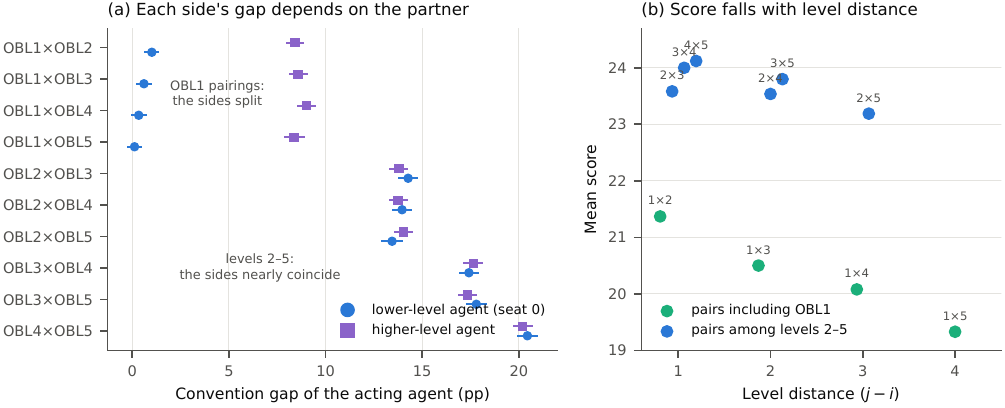}
  \caption{\textbf{Per-agent convention gap and score in the ten mixed-level
  pairings.} (a)~Convention gap of each acting agent per pairing; circles are the
  lower-level agent (always in seat~0), squares the higher-level one, with 95\%
  game-clustered bootstrap intervals. In pairings involving level~1 the two sides
  separate; among levels 2--5 they nearly coincide. (b)~Mean score of each pairing (no intervals shown) against level distance $j-i$, for levels $i < j$. Pairs including
  level~1 fall steeply with distance; pairs among levels 2--5 do not.}
  \label{fig:crosslevel}
\end{figure*}

The self-play ladder rises with level. It does not say what happens
when two levels meet. We played all ten mixed-level pairings ($1000$ games each,
replay-verified) and computed $\Delta$ separately for each side's own plays
(Fig.~\ref{fig:crosslevel}). In every game of this design the lower level held seat~0 and
moved first, and this analysis uses no seat-reversed games. Each side's statistics, and
their comparison with self-play (which pools both seats), therefore include any seat
effect. For scale, in the level-5 self-play baseline of \S\ref{sec:worth}, seat~0 has a
gap of $\tEightPZeroMaskedPlayerGapBase\pp$ and seat~1 $\tEightPZeroUnmaskedPlayerGapBase\pp$.
The previous study reports level~1's own gaps in these games \citep{paper1}; the analysis
below (the per-agent split across all ten pairings, its localization, and its seed
behavior) is new here.

\paragraph{The gap an agent realizes depends on its partner.}
Paired with a higher level, level~1's own aggregate gap falls toward zero: from
$\tFourgapLoOnexTwo\pp$ against level~2 to $\tFourgapLoOnexFive\pp$ against level~5, an
interval that includes zero. A near-zero aggregate can still contain plays that depart
from the literal prediction: on cards level~1 plays immediately after level~5
color-hinted them, its gap is $\tSixFreshGapOBLOnexOBLFiveLow\pp$ (\S\ref{app:mixed:cue}).
The higher-level agent, whose policy is unchanged, has gaps of $\tFourgapHiOnexTwo$ to
$\tFourgapHiOnexFour\pp$ across the four level-1 pairings, against its own self-play
values of $\gapOBLTwo$ to $\gapOBLFive\pp$. One reading of the higher agent's positive
gap is that level~1's hints tend to mark playable cards, so its hint choices carry some
information beyond their literal content. The partner dependence is not confined to one phase of the game: level~5's entire time
course flattens with level distance, its mid-game peak falling from
$\tcfigPartnerPeakOblFiveSelf\pp$ in self-play to $\tcfigPartnerPeakOblFiveWithOblThree$
with level~3 and $\tcfigPartnerPeakOblFiveWithOblOne\pp$ with level~1
(\S\ref{app:mixed:cue}, Fig.~\ref{fig:partnertc}).

\paragraph{The higher agent's loss rate rises, and score falls.}
The higher-level agent appears to keep taking convention-style plays that a level-1
partner's hints do not license, and its added failures are localized. On plays made immediately after a
partner color hint, the situation \S\ref{sec:when} identifies as the OBL hierarchy's
dominant cue, the higher agent's failure rate rises from
\dotrange{\tNineSelfplayFreshLossRangeLTwotoFive}\% in self-play to
$\tSixFreshLossOBLOnexOBLTwoHigh\%$ (level~2) through
$\tSixFreshLossOBLOnexOBLFiveHigh\%$ (level~5) when the hint comes from level~1. Level~1's
own failure rate in the same situation is \dotrange{\tNineOBLOneFreshLossRangeWithHigher}\%. Score
falls monotonically with level distance, from $\tFourscoreOnexTwo$ for $1\times2$ to
$\tFourscoreOnexFive$ for $1\times5$ (Fig.~\ref{fig:crosslevel}b). From $1\times3$ on,
the pair scores below level~1's own self-play value of $\scoreOBLOne$: level~1 scores
more with a copy of itself than with level~3, 4 or 5.

\paragraph{Levels 2--5 are mutually compatible and blend.}
In this seating, the two sides' gaps lie close together in every pairing among
levels 2--5, although the two levels differ in self-play. For $2\times5$, for instance,
the sides have gaps of $\tFourgapLoTwoxFive$ and $\tFourgapHiTwoxFive\pp$, against
self-play values of $\gapOBLTwoPt$ and $\gapOBLFivePt\pp$. The common value sits
between them, nearer the lower level, and score lands near the lower level's self-play
score. This is consistent with each agent taking mainly the plays its partner's hints
license. It is also what one expects if level $k$'s conventions are approximately a
subset of level $k+1$'s, rather than each level developing a separate system of its own.

\paragraph{Both patterns recur across seed combinations.}
The level-1 split appears in all $\tSevenControlAllPositive$ seed combinations of
$1\times5$, and among levels 2--5 the mean side difference over seed combinations stays
within $\pm\tSevenBlendTwotoFiveMaxAbsDelta\pp$ (Appendix~\ref{app:seeds}).

\subsection{The mixed-level cue map}
\label{app:mixed:cue}

\begin{figure}[H]
  \centering
  \includegraphics[width=\columnwidth]{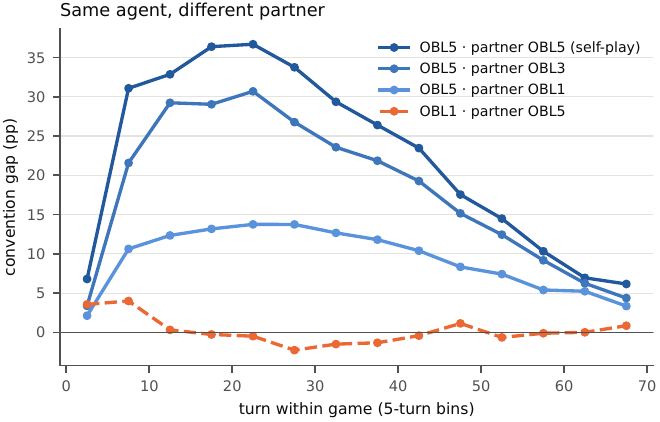}
  \caption{\textbf{Gap time course of a fixed agent under different partners.}
  Convention gap of one agent's own plays by 5-turn bin (bins with $\geq 250$ plays): level~5 in
  self-play, with level~3, and with level~1, and level~1 with level~5 (in mixed games the
  lower level held seat~0; self-play pools both seats). Point estimates; no intervals are
  shown. The whole time
  course, not one phase of it, flattens with level distance.}
  \label{fig:partnertc}
\end{figure}

The per-agent time courses under different partners appear in
Fig.~\ref{fig:partnertc}; the cue-level view localizes them. On plays made immediately
after a partner color hint, the agent facing a level-1 partner keeps entering the
situation ($\tSixFreshShareOBLOnexOBLTwoHigh$ to $\tSixFreshShareOBLOnexOBLFiveHigh\%$ of
its plays) while its loss rate there rises with its own level, from
$\tSixFreshLossOBLOnexOBLTwoHigh\%$ (level~2) to $\tSixFreshLossOBLOnexOBLFiveHigh\%$
(level~5), against \dotrange{\tNineSelfplayFreshLossRangeLTwotoFive}\% in self-play, and its
cue gap falls to $\tSixFreshGapOBLOnexOBLFiveHighPt\pp$ from a self-play value near
$\tFiveGapFreshColorOBLFiveTrend\pp$. Level~1 fails rarely in the same situation
(loss \dotrange{\tNineOBLOneFreshLossRangeWithHigher}\%; gap
$\tSixFreshGapOBLOnexOBLTwoLowPt$ to $\tSixFreshGapOBLOnexOBLFourLowPt\pp$). Between
levels 2--5 the cue blends unevenly: the lower
partner's cue gap rises above its own self-play value (level~2 with level~5:
$\tSixFreshGapOBLTwoxOBLFiveLowPt\pp$) at a loss rate that stays low
($\tSixFreshLossOBLTwoxOBLFiveLow\%$), while the higher partner's falls toward the lower partner's
value ($\tSixFreshGapOBLTwoxOBLFiveHighPt\pp$). Both sides' overall gaps meet near the
blended values of \S\ref{sec:who}.

%% file: appendix/C_seeds.tex
\section{Seed robustness}
\label{app:seeds}

\subsection{The seed sweep}

Appendix~\ref{app:mixed} reports single-seed pairings: one released model per level.
This appendix asks whether its two results, the level-1 split and the blend among
levels 2--5, depend on which seed models were paired.
To test this, every
pairing among levels 2--5, plus the level-$1{\times}5$ control, was played for all $25$
seed combinations (seeds a--e on each side, $200$ games per combination;
$\tSevenGamesTotal$ games in total, each replay-verified). The unit of analysis is the
seed combination: the primary statistic per level pair is the across-combination mean of
the side difference $\delta = \Delta_{\text{higher}} - \Delta_{\text{lower}}$, with a
$t$-based $95\%$ interval over the $25$ combinations
(Table~\ref{tab:seedsweep}). The combinations share seed models on each side and are
not independent replications, so the $t$ interval may be too narrow for inference
beyond these models. As in Appendix~\ref{app:mixed}, the lower level held seat~0 in
every game.

\input{appendix/tables/tabB_seeds}

Both patterns recur across seed combinations. Among levels 2--5 the mean
$\delta$ stays within $\pm\tSevenBlendTwotoFiveMaxAbsDelta\pp$ of zero for every pairing,
an order of magnitude below the control's $\tSevenOBLOnexOBLFiveDelta\pp$, which is
positive in all $\tSevenControlAllPositive$ combinations (per-combination range
$\tSevenControlDeltaMin$ to $\tSevenControlDeltaMax\pp$). The sweep also points to two
asymmetries that the single-seed analysis did not show: $\delta$ is positive, with an
interval above zero, for $2{\times}4$ ($\tSevenOBLTwoxOBLFourDelta\pp$) and for
$2{\times}5$ ($\tSevenOBLTwoxOBLFiveDelta\pp$), so when level~2 is the lower partner the
higher side's gap may be slightly larger. With seats fixed, these differences cannot be
separated from seat: in level-5 self-play the two seats' gaps are
$\tEightPZeroMaskedPlayerGapBase$ and $\tEightPZeroUnmaskedPlayerGapBase\pp$
(Appendix~\ref{app:mixed}). The $4{\times}5$ interval touches zero from below
($\tSevenOBLFourxOBLFiveDelta\pp$); given the shared models and the number of
intervals, we do not read it as an asymmetry. The last column estimates the share of across-combination variance in
$\delta$ not explained by game-sampling noise; it rests on bootstrap standard errors and
is approximate.

\subsection{Self-play across seeds}

Self-play gaps vary across the five seed models with standard deviations of
$\tSevenSelfplaySdOBLTwo\pp$ (level~2), $\tSevenSelfplaySdOBLThree$ (level~3),
$\tSevenSelfplaySdOBLFour$ (level~4) and $\tSevenSelfplaySdOBLFive\pp$ (level~5).
The spread is larger at higher levels. Per-level ranges overlap across seeds, but the means
over seeds rise from $\tSevenSelfplayMeanOBLTwo$ (level~2) to $\tSevenSelfplayMeanOBLFive\pp$
(level~5), and the ordering $\Delta_{2} < \Delta_{3} < \Delta_{4} < \Delta_{5}$ holds
\emph{within} each of the \tSevenMonotoneSeeds{} seeds when models are matched by seed
letter.

\subsection{Seed heterogeneity sits in the taxonomy tail}

Within the induced taxonomy of \S\ref{sec:where}, per-seed cell values were checked for
the top-ten categories by absolute gap at each of levels 2--5 ($\txHetChecked$ cells;
$\txHetFlags$ flagged with across-seed excess variance beyond game-sampling noise). Head
cells are seed-stable in relative terms: their excess standard deviations are
\txHetHeadExcessRange{}. Every large-excess flag sits in a tail cell
holding at most \txHetLargeExcessMaxShare{} of plays: the widest are a
discard-context color cell at level~3 (\txHetOBLThreeDiscardC\pp{}), an unhinted
cell at level~4 (\txHetOBLFourNeverHinted\pp{}), and a repeat-color-hint cell at
level~5 (\txHetOBLFiveCCplusOlder\pp{}); each parenthesis gives the range across seeds
and the excess standard deviation. Different training runs
agree on the head of the taxonomy and differ in its tail.

%% file: appendix/tables/tabB_seeds.tex
\begin{table*}[t]
  \centering\footnotesize
  \caption{Seed sweep: all $25$ seed combinations per level pair, $200$ games each.
  Gaps in pp (mean $\pm$ sd across combinations); $\delta$ is the across-combination
  mean of $\Delta_{\text{higher}} - \Delta_{\text{lower}}$ with a $t$-based $95\%$
  interval treating combinations as the unit (they share seed models on each side, so
  the interval may be too narrow); the last column is the estimated share
  of across-combination variance in $\delta$ beyond game-sampling noise (approximate).}
  \label{tab:seedsweep}
  \setlength{\tabcolsep}{4pt}
  \begin{tabular}{lccccc}
  \toprule
  Pair & Score & $\Delta$, lower & $\Delta$, higher & $\delta$ [95\% CI] & Seed share (\%) \\
  \midrule
  $2\times3$ & $\tSevenOBLTwoxOBLThreeScore \pm \tSevenOBLTwoxOBLThreeScoreSd$ & $\tSevenOBLTwoxOBLThreeGapLow \pm \tSevenOBLTwoxOBLThreeGapLowSd$ & $\tSevenOBLTwoxOBLThreeGapHigh \pm \tSevenOBLTwoxOBLThreeGapHighSd$ & $\tSevenOBLTwoxOBLThreeDeltaPt~[\tSevenOBLTwoxOBLThreeDeltaCiLo, \tSevenOBLTwoxOBLThreeDeltaCiHi]$ & $\tSevenOBLTwoxOBLThreeSeedShare$ \\
  $2\times4$ & $\tSevenOBLTwoxOBLFourScore \pm \tSevenOBLTwoxOBLFourScoreSd$ & $\tSevenOBLTwoxOBLFourGapLow \pm \tSevenOBLTwoxOBLFourGapLowSd$ & $\tSevenOBLTwoxOBLFourGapHigh \pm \tSevenOBLTwoxOBLFourGapHighSd$ & $\tSevenOBLTwoxOBLFourDeltaPt~[\tSevenOBLTwoxOBLFourDeltaCiLo, \tSevenOBLTwoxOBLFourDeltaCiHi]$ & $\tSevenOBLTwoxOBLFourSeedShare$ \\
  $2\times5$ & $\tSevenOBLTwoxOBLFiveScore \pm \tSevenOBLTwoxOBLFiveScoreSd$ & $\tSevenOBLTwoxOBLFiveGapLow \pm \tSevenOBLTwoxOBLFiveGapLowSd$ & $\tSevenOBLTwoxOBLFiveGapHigh \pm \tSevenOBLTwoxOBLFiveGapHighSd$ & $\tSevenOBLTwoxOBLFiveDeltaPt~[\tSevenOBLTwoxOBLFiveDeltaCiLo, \tSevenOBLTwoxOBLFiveDeltaCiHi]$ & $\tSevenOBLTwoxOBLFiveSeedShare$ \\
  $3\times4$ & $\tSevenOBLThreexOBLFourScore \pm \tSevenOBLThreexOBLFourScoreSd$ & $\tSevenOBLThreexOBLFourGapLow \pm \tSevenOBLThreexOBLFourGapLowSd$ & $\tSevenOBLThreexOBLFourGapHigh \pm \tSevenOBLThreexOBLFourGapHighSd$ & $\tSevenOBLThreexOBLFourDeltaPt~[\tSevenOBLThreexOBLFourDeltaCiLo, \tSevenOBLThreexOBLFourDeltaCiHi]$ & $\tSevenOBLThreexOBLFourSeedShare$ \\
  $3\times5$ & $\tSevenOBLThreexOBLFiveScore \pm \tSevenOBLThreexOBLFiveScoreSd$ & $\tSevenOBLThreexOBLFiveGapLow \pm \tSevenOBLThreexOBLFiveGapLowSd$ & $\tSevenOBLThreexOBLFiveGapHigh \pm \tSevenOBLThreexOBLFiveGapHighSd$ & $\tSevenOBLThreexOBLFiveDeltaPt~[\tSevenOBLThreexOBLFiveDeltaCiLo, \tSevenOBLThreexOBLFiveDeltaCiHi]$ & $\tSevenOBLThreexOBLFiveSeedShare$ \\
  $4\times5$ & $\tSevenOBLFourxOBLFiveScore \pm \tSevenOBLFourxOBLFiveScoreSd$ & $\tSevenOBLFourxOBLFiveGapLow \pm \tSevenOBLFourxOBLFiveGapLowSd$ & $\tSevenOBLFourxOBLFiveGapHigh \pm \tSevenOBLFourxOBLFiveGapHighSd$ & $\tSevenOBLFourxOBLFiveDeltaPt~[\tSevenOBLFourxOBLFiveDeltaCiLo, \tSevenOBLFourxOBLFiveDeltaCiHi]$ & $\tSevenOBLFourxOBLFiveSeedShare$ \\
  $1\times5$ (control) & $\tSevenOBLOnexOBLFiveScore \pm \tSevenOBLOnexOBLFiveScoreSd$ & $\tSevenOBLOnexOBLFiveGapLow \pm \tSevenOBLOnexOBLFiveGapLowSd$ & $\tSevenOBLOnexOBLFiveGapHigh \pm \tSevenOBLOnexOBLFiveGapHighSd$ & $\tSevenOBLOnexOBLFiveDeltaPt~[\tSevenOBLOnexOBLFiveDeltaCiLo, \tSevenOBLOnexOBLFiveDeltaCiHi]$ & $\tSevenOBLOnexOBLFiveSeedShare$ \\
  \bottomrule
  \end{tabular}
\end{table*}

%% file: appendix/C_interventions.tex
\section{Intervention details}
\label{app:interventions}

This appendix gives the details behind the intervention results of \S\ref{sec:worth}:
the verification gates, the full condition table, the replication of the one anomalous
mask, and a category-level check of the rerouting. Apart from the level-2 contrast, the
masks act on the seed-a model of level~5, which is in the released archive but not among
the results published with OBL \citep{hu2021obl} (\S\ref{sec:hierarchy}); the replication
adds level-5 seed~b and level-4
seed~a. The level-5 baseline uses this model on its own environment seeds, and its gap
($\eFiveBaselineGap\pp$) differs from the ladder value ($\gapOBLFivePt\pp$).

\subsection{Verification gates}

Two gates precede every full run. First, the category that the masking code assigns to
each play during the game must match the category that the analysis assigns afterwards;
on the baseline games the two agree on every play a mask would block
($\eFiveGateDescriptorAgreement$). Second, the masking code with no mask active must
reproduce the baseline games: $\eFiveGateLogOnlyIdentical$ games match action for action,
deck and score. All mask conditions share environment
seeds with the baseline, and the full $50$-card decks are identical per seed across every
condition (\eFiveDeckIdentity{}), so score differences are paired per deck. Every
condition replays to its ground-truth score.

\subsection{The condition table}

Targets are the top eight categories by observational contribution (share $\times$ gap,
estimation half) among categories holding at least $1\%$ of plays, three low-gap controls
(two fully identified cells, whose gap is zero by construction, and the large low-gap rank
cell at lag $4{+}$), and cumulative
masks of the top $1/2/4/6/8$ (top-1 coincides with single \#1). Frequency-matched random
controls suppress the greedy play with probability matched to the trigger rates of
targets \#1, \#4 and \#8 ($\eFiveControlRates$ triggers per game) and define the null
cost curve: piecewise linear through the origin and the three controls in realized
trigger rate, linearly extrapolated beyond the largest control rate, and re-fit inside
every bootstrap draw so the excess interval carries the control curve's own uncertainty.
Rows whose realized trigger rate exceeds the largest control rate (the cumulative
top-2, top-4, top-6 and top-8 masks) rely on that extrapolation and are marked $^\dagger$ in
Table~\ref{tab:conditions}; the
super-additivity claim in \S\ref{sec:worth} uses raw paired costs and does not depend
on it.

\input{appendix/tables/tabC_conditions}

The three low-gap controls (Table~\ref{tab:conditions}) show that the control curve is
accurate to about a tenth of a point (\maintext \S\ref{sec:worth}).

\paragraph{Why no dose--response fit is reported.}
An ordinary least-squares fit of the eight single-mask excess costs on contribution gives
a slope of $\eFiveSlopeContrib$ points per unit of contribution (contribution as a
probability; interval from the game-level bootstrap) with $R^2 = \eFiveRsqContrib$;
on the category's gap alone $R^2 = \eFiveRsqGap$, on its share alone
$\eFiveRsqShare$. We do not interpret these fits. They rest on eight cells, six of which
sit within about one percentage point of contribution of each other, so categories~\#1 and~\#2
carry most of the leverage, and neither of their excess intervals excludes zero. The fit
is also sensitive to category~\#6, whose replication was inconclusive (\S\ref{app:cat6}). The
bootstrap interval treats the eight categories as fixed and therefore says nothing
about how excess cost would vary across other categories.

\subsection{Redistribution under the masks}

Masking a color cell delays some of its plays. Under the
top-1 mask the delayed color-only category (lag $2$--$3$) grows from
$\eFiveConlyLagTwoThreeShareBase\%$ of plays to $\eFiveConlyLagTwoThreeShareTopOne\%$ at
a gap of $\eFiveConlyLagTwoThreeGapTopOne\pp$; under the top-8 mask it reaches
$\eFiveConlyLagTwoThreeShareTopEight\%$ at $\eFiveConlyLagTwoThreeGapTopEight\pp$. The
total gap falls from $\eFiveBaselineGap$ to $\tOneFourTotalGapCumEight\pp$ while score
falls $\eFiveCumTopEightCost$ points.
The single-mask and cumulative results are plotted in \maintext
Fig.~\ref{fig:intervention}b and~c.

\subsection{Replication of the category-6 anomaly}
\label{app:cat6}

The apparent score \emph{improvement} from masking category~\#6 was re-tested out of sample
before being interpreted: fresh decks disjoint from every prior condition, three agents
(level-5 seeds a and b, level-4 seed a), each with its own baseline and frequency-matched
random control, and a decision rule fixed in advance (replicates if all three excesses
are positive and at least two per-agent intervals exclude zero; fails if the pooled
interval contains zero). Outcome: \MakeLowercase{\eFivebDecision}. The excesses are positive for all
three agents ($\eFivebExcessOBLFiveA$, $\eFivebExcessOBLFiveB$, $\eFivebExcessOBLFourA$)
and the inverse-variance pooled excess is $\eFivebPooledExcess$, but only
$\eFivebNumAgentCIsExcludeZero$ per-agent interval excludes zero, and the raw score gain
in Table~\ref{tab:conditions} ($\tOneFourCostSingleSixPt$ points) did not reproduce. The
supported reading is that this convention carries no detectable positive value
(``cheaper than random disruption,'' not beneficial), and that the first estimate was
inflated by selecting among eight categories. The substitution pattern does reproduce: under
the mask, \eFivebRankHintSubstitutionRange{} of masked plays substitute into another rank
hint, as in the original run.

\subsection{Taxonomy validation on the intervention games}

Applying the frozen taxonomy to the masked games checks the rerouting at category
level, independently of the score analysis. Under the immediacy mask, level-5's
color-only lag-1 categories empty by construction
(\txMaskBaseColorLagOneShare{} of baseline plays at $\txMaskBaseColorLagOneGap\pp$) and
many displaced plays reappear as color-only lag-$2$--$3$ plays at nearly undiminished
gap: that category's share rises from \txMaskLagTwoThreeShareBase{} to
\txMaskLagTwoThreeShareImm{} at $\txMaskLagTwoThreeGapImm\pp$, with a smaller spill
into lag $4{+}$ (\txMaskLagFourShareBase{} $\rightarrow$ \txMaskLagFourShareImm{} at
$\txMaskLagFourGapImm\pp$). Under the persistent mask the color-only categories stay
empty (\txMaskPersistColorShare{}) and play shifts onto fully identified cards, whose gap
is zero by construction (\txMaskFullyTypedBase{} $\rightarrow$ \txMaskFullyTypedPersist{}
of plays). Level~2
reroutes less. Its delayed color-only gap at baseline is
$\txMaskOBLTwoDelayedGapBase\pp$, against level-5's $\txMaskOBLFiveDelayedGapBase\pp$, and
under the immediacy mask its color-only lag-$2$--$3$ share reaches
\txMaskOBLTwoLagTwoThreeShareImm{}, against level-5's \txMaskLagTwoThreeShareImm{}.

%% file: appendix/tables/tabC_conditions.tex
\begin{table*}[t]
  \centering\footnotesize
  \caption{All category-mask conditions ($1000$ paired games each). Contribution is the
  observational share $\times$ gap of the target set (estimation half, pp); cost is the
  paired per-deck score change against the shared baseline (points; negative values are
  losses); excess is the cost minus the frequency-matched random-mask cost at the same
  realized trigger rate, with the control curve re-fit inside every bootstrap draw, so a
  positive excess means the mask costs less than a random mask; total gap is the corpus
  gap under the mask (pp). Frequency-matched random controls define the zero line and are omitted
  (excess $= 0$ by construction). $^\dagger$The excess relies on extrapolating the
  control curve beyond the largest control rate.}
  \label{tab:conditions}
  \setlength{\tabcolsep}{4pt}
  \begin{tabular}{lcccc}
  \toprule
  Condition & Contribution & Cost & Excess [95\% CI] & Total gap \\
  \midrule
  single \#1 & $\tOneFourContribSingleOne$ & $\tOneFourCostSingleOnePt$ & $\tOneFourExcessSingleOnePt~\tOneFourExcessCISingleOne$ & $\tOneFourTotalGapSingleOnePt$ \\
  single \#2 & $\tOneFourContribSingleTwo$ & $\tOneFourCostSingleTwoPt$ & $\tOneFourExcessSingleTwoPt~\tOneFourExcessCISingleTwo$ & $\tOneFourTotalGapSingleTwoPt$ \\
  single \#3 & $\tOneFourContribSingleThree$ & $\tOneFourCostSingleThreePt$ & $\tOneFourExcessSingleThreePt~\tOneFourExcessCISingleThree$ & $\tOneFourTotalGapSingleThreePt$ \\
  single \#4 & $\tOneFourContribSingleFour$ & $\tOneFourCostSingleFourPt$ & $\tOneFourExcessSingleFourPt~\tOneFourExcessCISingleFour$ & $\tOneFourTotalGapSingleFourPt$ \\
  single \#5 & $\tOneFourContribSingleFive$ & $\tOneFourCostSingleFivePt$ & $\tOneFourExcessSingleFivePt~\tOneFourExcessCISingleFive$ & $\tOneFourTotalGapSingleFivePt$ \\
  single \#6 & $\tOneFourContribSingleSix$ & $\tOneFourCostSingleSixPt$ & $\tOneFourExcessSingleSixPt~\tOneFourExcessCISingleSix$ & $\tOneFourTotalGapSingleSixPt$ \\
  single \#7 & $\tOneFourContribSingleSeven$ & $\tOneFourCostSingleSevenPt$ & $\tOneFourExcessSingleSevenPt~\tOneFourExcessCISingleSeven$ & $\tOneFourTotalGapSingleSevenPt$ \\
  single \#8 & $\tOneFourContribSingleEight$ & $\tOneFourCostSingleEightPt$ & $\tOneFourExcessSingleEightPt~\tOneFourExcessCISingleEight$ & $\tOneFourTotalGapSingleEightPt$ \\
  control A (fully identified) & $\tOneFourContribCtrlA$ & $\tOneFourCostCtrlAPt$ & $\tOneFourExcessCtrlAPt~\tOneFourExcessCICtrlA$ & $\tOneFourTotalGapCtrlAPt$ \\
  control B (fully identified) & $\tOneFourContribCtrlB$ & $\tOneFourCostCtrlBPt$ & $\tOneFourExcessCtrlBPt~\tOneFourExcessCICtrlB$ & $\tOneFourTotalGapCtrlBPt$ \\
  control C (rank, lag $4{+}$) & $\tOneFourContribCtrlRlagFour$ & $\tOneFourCostCtrlRlagFourPt$ & $\tOneFourExcessCtrlRlagFourPt~\tOneFourExcessCICtrlRlagFour$ & $\tOneFourTotalGapCtrlRlagFourPt$ \\
  cumulative top-2 & $\tOneFourContribCumTwo$ & $\tOneFourCostCumTwoPt$ & $\tOneFourExcessCumTwoPt~\tOneFourExcessCICumTwo^{\dagger}$ & $\tOneFourTotalGapCumTwoPt$ \\
  cumulative top-4 & $\tOneFourContribCumFour$ & $\tOneFourCostCumFourPt$ & $\tOneFourExcessCumFourPt~\tOneFourExcessCICumFour^{\dagger}$ & $\tOneFourTotalGapCumFourPt$ \\
  cumulative top-6 & $\tOneFourContribCumSix$ & $\tOneFourCostCumSixPt$ & $\tOneFourExcessCumSixPt~\tOneFourExcessCICumSix^{\dagger}$ & $\tOneFourTotalGapCumSixPt$ \\
  cumulative top-8 & $\tOneFourContribCumEight$ & $\tOneFourCostCumEightPt$ & $\tOneFourExcessCumEightPt~\tOneFourExcessCICumEight^{\dagger}$ & $\tOneFourTotalGapCumEightPt$ \\
  \bottomrule
  \end{tabular}
\end{table*}

%% file: appendix/D_corpus.tex
\section{Human corpus}
\label{app:corpus}

This appendix documents the two human corpora (provenance, filters, clustering and the
convention community they come from) and gives the full versions of the human results
summarized in \S\ref{sec:what}.

\subsection{Provenance and filters}

The corpus is drawn from public logs of two-player, no-variant games on the hanab.live
platform \citep{hanablive}, the platform the convention community plays on, with usernames replaced by salted hashes before any
analysis. Two filters define the analysis set. The completed-games filter removes games
whose logs end abnormally (timeouts, terminations, or logs that stop with cards in the
deck and no terminal state). The bot exclusion has two tiers: a game-level tier for
verified bot families matched by account-name pattern, and an account-level tier
removing plays by flagged accounts (accounts that play almost only with verified bots,
pairs of accounts with identical timing statistics, and accounts whose game notes carry a
bot's version tag; criteria as in the previous study \citep{paper1}, identified by hashed
id). The result is $\hhGamesTwoTier$ games,
$\hhPlaysTwoTier$ plays, $\hhPlayersTwoTier$ players and $\hhPairsTwoTier$ pairs. A
sensitivity corpus applying only the verified-bot tier holds $\hhGamesTierOne$ games /
$\hhPlaysTierOne$ plays and gives $\hhGapJointTierOne\pp$.

\subsection{Statements shared with the prior paper}

This corpus, its filters and its posteriors are those of the previous study that
introduced the convention gap \citep{paper1} (``the prior paper'' below).

\emph{Posterior variant.} Both papers report the joint (full-hand) posterior, which
gives a corpus headline of $\hhGapJointTwoTierPt\pp$. The per-card approximation gives
$\hhGapMarginalTwoTierPt\pp$ on the same corpus. The two variants are never mixed within
a figure or a comparison.

\emph{The Tier-1 filter, as implemented.} The Tier-1 filter removes every game in which
a verified bot account made a play. One game in which a verified bot only hinted and
discarded is therefore retained; excluding it as well gives $\poneTierOneStrictGames$
games and $\poneTierOneStrictPlays$ plays and a gap of $\poneTierOneStrictGap\pp$, against
$\hhGapJointTwoTierPt\pp$ \citep{paper1}.

\subsection{Corpus identity with the prior paper}

The identity is exact, not approximate, and was established by a three-part check:
(i)~the game sets are identical (sorted id lists, symmetric difference empty);
(ii)~the filter chain is the \emph{same code}, imported rather than reimplemented; and
(iii)~the joint posteriors agree play-for-play, bit-identically, over all
$\hhPlaysTwoTier$ plays against an independent run of the prior paper's engine on the
raw logs.

\subsection{Clustering: pair against game}

Pairs recur across games, so the primary human intervals cluster at the pair level
($\hhPairsTwoTier$ pairs); game-clustered intervals are narrower. Side by side, joint
posterior: corpus gap $\hhGapJointTwoTier$ (pair) against
$\hhGapJointTwoTierCiGame$ (game); Tier-1-only sensitivity $\hhGapJointTierOne$ (pair)
against $\hhGapJointTierOneCiGame$ (game). The documented exceptions are the
nearest-level bootstrap and the mixture fit of Appendix~\ref{app:depth}
(\S\ref{app:depth:nearest} and \S\ref{app:depth:fit}), which are game-clustered by design,
for comparability with the agent corpora; their pair-clustered sensitivity is in
\S\ref{app:depth:paircl}.

\subsection{The convention community}
\label{app:community}

The platform's community publishes and teaches a convention system \citep{hgroup}: play hints and the focus rule, chop and save conventions, and the finesse family are written down rather than negotiated at the table. The human corpus therefore describes \emph{taught} play, not typical human play. Pair experience, counted within the corpus, moves the gap little: stratifying pairs by games played together gives $\hhExpOneGameGap\pp$ (one game together, $\hhExpOneGamePairs$ pairs), $\hhExpTwoToFourGap\pp$ (2--4 games), and $\hhExpFivePlusGap\pp$ (five or more), and near-strangers (one game together, both players in at most two corpus games) reach $\hhExpNearStrangersGap\pp$. This is consistent with players who arrive already sharing a taught system.

\subsection{The AH2AC2 corpus}
\label{app:ah2ac2}

The replication corpus (\S\ref{sec:replication}) is processed identically from the
open release: conversion to the common game format, replay verification against the
released scores ($\axNGamesReleased$ of $\axNGamesReleased$ games reproduce exactly),
then the same completed-games criterion, which removes $\axDroppedStuckGames$ games.
Every one is a verifiable ``no further points possible'' early ending, a completed
game our criterion nonetheless excludes; they are retained as a sensitivity
($\axGapJointFullOneEightFiveEight\pp$ on the full release against $\axGapJoint\pp$ on
the filtered corpus). The open data carry no player identifiers, so the bot exclusion
cannot be applied and all AH2AC2 intervals cluster by game. Score selection in that
collection (minimum recorded score $\axScoreFloorMin$; no strikeouts) is consistent with
its higher aggregate (\S\ref{sec:replication}); restricting the hanab.live corpus to
games at or above the same floor gives $\hhScoreFloorGap\pp$ over $\hhScoreFloorGames$
games.

\subsection{The shared taxonomy ranked by the human--level-5 difference}
\label{app:shareddiff}

Ranking the cells of the shared taxonomy (\S\ref{sec:where}) by the absolute
difference between the human and level-5 gaps (Fig.~\ref{fig:shareddiff}) ignores how often each cell occurs. The split in
\S\ref{sec:humans} is an accounting identity and assigns no cause. Its within-cell term
weights each cell's gap difference by the human share of plays. Its other term weights
each cell's share difference by the level-5 gap.
The three largest differences are the two indirect-reference cells and a delayed color
cell. Level~5's gap is higher on three newest-card cells played right after a hint and,
by small amounts, on three merged cells; the cells that are zero by construction close
the list at exactly zero. The two indirect-reference cells are merged residual cells,
which the next subsection resolves.

\ifdefined\appendixCarriesMovedFigs
\begin{figure*}[tp]
  \centering
  \includegraphics[width=\textwidth,height=0.8\textheight,keepaspectratio]{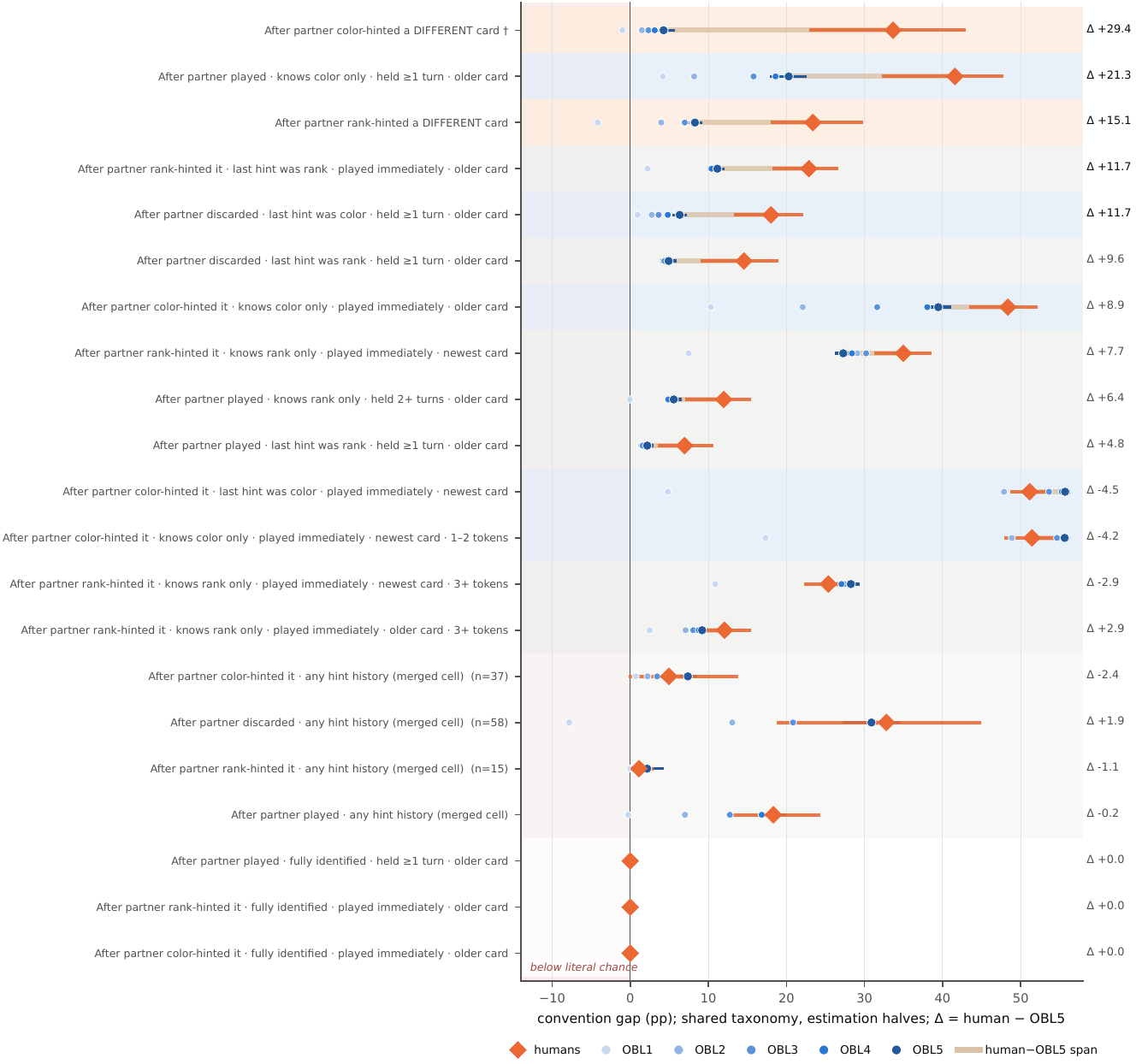}
  \caption{\textbf{Shared-taxonomy cells ranked by the absolute human$\,$--$\,$level-5
  difference.} Same cells and basis as \maintext Fig.~\ref{fig:shared}, whose caption
  gives the key to the row labels, $\dagger$ and $n$; $\Delta$ is the human gap minus the
  level-5 gap (pp). Diamonds: humans, with 95\% pair-clustered bootstrap intervals; dots:
  levels 1--5 (darker is higher); the tan bar spans the human and level-5 values; the
  strip left of zero marks gaps below literal chance. Row shading: indirect reference
  (orange), color known (blue), rank known (gray), merged (light gray), fully identified
  (unshaded). The three largest differences are the two indirect-reference cells and a
  delayed color cell.}
  \label{fig:shareddiff}
\end{figure*}
\fi

\subsection{The indirect-reference category, decomposed}

\Maintext Table~\ref{tab:usage}'s indirect-reference rows are merged cells, defined only
by the partner's preceding action, and their contents differ across populations
(Fig.~\ref{fig:composition}). The numbers in this paragraph use the frozen level-5
taxonomy on the full corpus, the basis of the right-hand panels of
Fig.~\ref{fig:composition}; on that basis the human
color cell holds $\hhCatIndirectColorN$ plays. Decomposed by the played card's own
hint history, it is $\tTwelveANeverHintedShare\%$ unhinted cards (the finesse
configuration, $\tTwelveANeverHintedGap\pp$ over $\tTwelveANeverHintedN$ plays),
$\tTwelveAColorOnlyShare\%$ color-known cards ($\tTwelveAColorOnlyGap\pp$), and a
remainder too small to estimate. The left-hand panels of Fig.~\ref{fig:composition}
use a different basis, the shared taxonomy on estimation halves, whose residual cell
is broader because it must be supported by all six populations; its unhinted share
is therefore lower than the frozen-cell figure quoted here, and the two panels are not
meant to match. Table~\ref{tab:silencesplit} gives the reverse decomposition: all unhinted plays split by the partner's preceding action. In human play, $\silHHAllHintOtherN$ of the $\silHHAllAllN$ unhinted plays follow a hint to another card (gap $\silHHAllHintOtherGap\pp$), against $\silOBLFiveAllHintOtherN$ of $\silOBLFiveAllAllN$ for level~5. On the remaining unhinted plays, humans and level~5 differ by $\silDiffHHMinusOBLFiveAllClean\pp$; in AH2AC2, which keeps no struck-out games, the difference is $\silDiffAXMinusOBLFiveAllClean\pp$. The color-known sub-cell is the composition-matched
comparison used in \S\ref{sec:replication}: $\axColorKnownSubcellGap\pp$ in AH2AC2 against $\tTwelveAColorOnlyGap\pp$ here.

\begin{table*}[tp]
  \centering\footnotesize
  \caption{\textbf{Unhinted plays by the partner's preceding action.} Gap (pp) with
  95\% interval; number of plays in parentheses. ``Other'' is a partner play or
  discard, or the first move. Humans: pair-clustered; AH2AC2 and level~5 (seed~a): game-clustered. In the hanab.live corpus, the $\silHHAllColorOtherN$ unhinted plays that follow a color hint to another card (the finesse configuration) have a mean literal failure probability of $\silHHAllColorOtherPost\%$ and a failure rate of $\silHHAllColorOtherLoss\%$.}
  \label{tab:silencesplit}
  \begin{tabular}{lccc}
  \toprule
  Partner's preceding action & Humans (hanab.live) & Humans (AH2AC2) & Level~5 \\
  \midrule
  Hint to another card & $\silHHAllHintOtherGap$ ($\silHHAllHintOtherN$) & $\silAXAllHintOtherGap$ ($\silAXAllHintOtherN$) & $\silOBLFiveAllHintOtherGap$ ($\silOBLFiveAllHintOtherN$) \\
  Other & $\silHHAllCleanGap$ ($\silHHAllCleanN$) & $\silAXAllCleanGap$ ($\silAXAllCleanN$) & $\silOBLFiveAllCleanGap$ ($\silOBLFiveAllCleanN$) \\
  All unhinted plays & $\silHHAllAllGap$ ($\silHHAllAllN$) & $\silAXAllAllGap$ ($\silAXAllAllN$) & $\silOBLFiveAllAllGap$ ($\silOBLFiveAllAllN$) \\
  \midrule
  Difference from level~5, other & $\silDiffHHMinusOBLFiveAllClean$ & $\silDiffAXMinusOBLFiveAllClean$ & \\
  \bottomrule
  \end{tabular}
\end{table*}

\ifdefined\appendixCarriesMovedFigs
\begin{figure*}[tp]
  \centering
  \includegraphics[width=\textwidth,height=0.8\textheight,keepaspectratio]{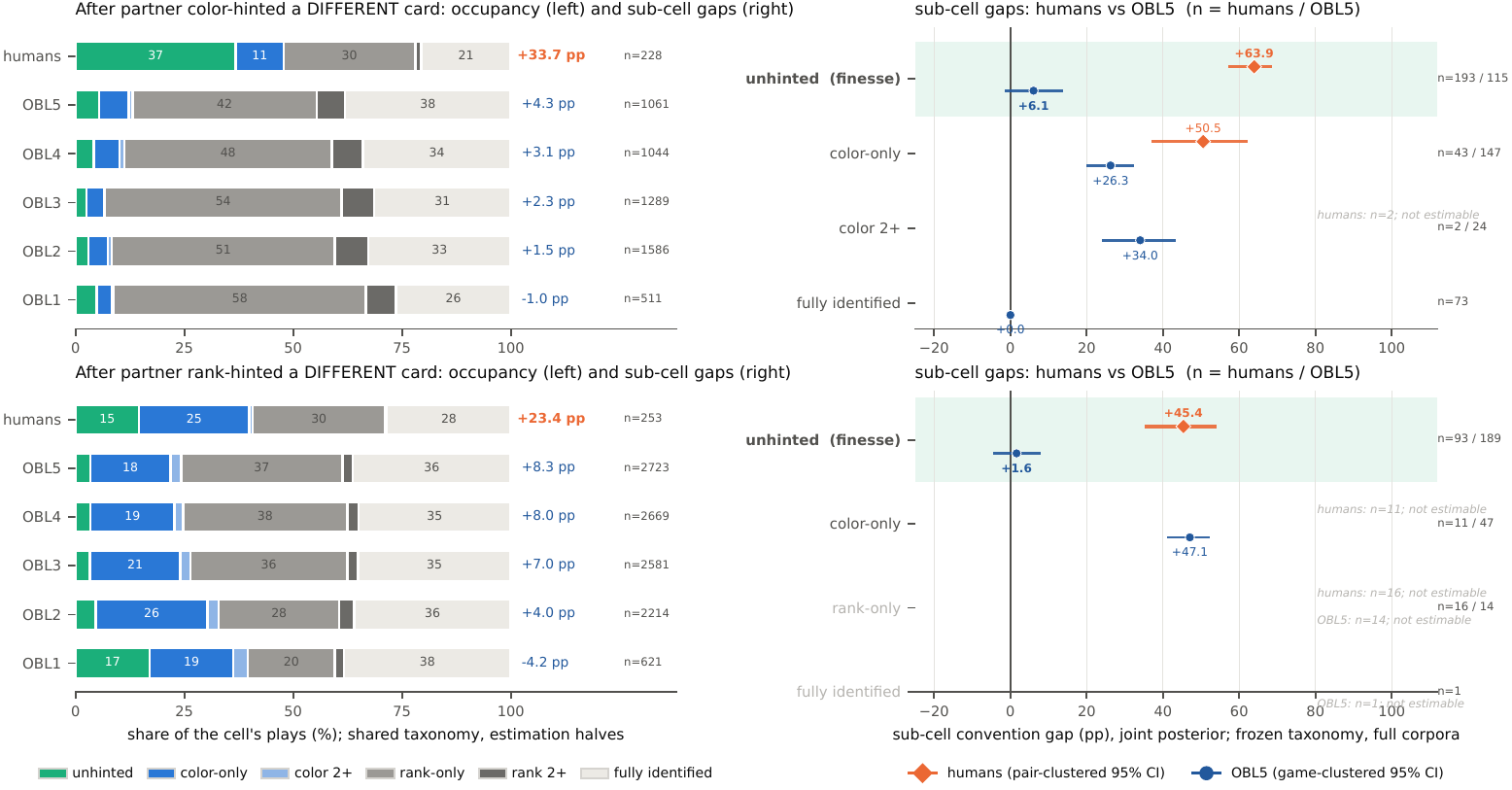}
  \caption{\textbf{The two indirect-reference residual cells resolved by the played
  card's own hint history.} Left: sub-cell occupancy per population (shared taxonomy,
  estimation halves): human entries are dominated by unhinted finesse targets, agent
  entries by rank-known and fully-identified cards. The value right of each bar is that
  population's gap in the whole cell, with its number of plays. Right: human and level-5
  gaps per
  sub-cell (frozen level-5 taxonomy, full corpora, a narrower cell than on the left, so
  the two panels' shares differ by construction; humans pair-clustered, level~5
  game-clustered; sub-cells with $n<20$ are marked not estimable rather than zero).
  The highlighted finesse rows show the largest differences: humans
  $\tTwelveANeverHintedGap$ and $\fcRnkHumanUnhintedGapPt\pp$ where level~5 reads
  $\fcColOblFiveUnhintedGapPt$ and $\fcRnkOblFiveUnhintedGapPt\pp$. On every sub-cell
  estimable for both populations, the human value exceeds level~5's, with disjoint
  intervals.}
  \label{fig:composition}
\end{figure*}
\fi

\subsection{Replication on the AH2AC2 corpus in full}

The corpus above reflects one set of collection and filtering choices. The AH2AC2 benchmark
\citep{ah2ac2} independently released $\axNGamesReleased$ two-player games collected from the same
platform and convention community. The two collections share \emph{no} game records: content identity over deck and
full action sequence matches $0$ of $\axNGamesReleased$ against the raw hanab.live logs used here. This
is therefore an independent collection from the same population (deck-level
coincidences exist, but only because the platform deals seeded decks to many
tables).
Player overlap is unknowable, because the open data carry no player identifiers
(\S\ref{app:ah2ac2}).

Replaying all games through the identical pipeline (replay-verified against the released
scores, $\axNGamesReleased/\axNGamesReleased$; $\axNGames$ games after the completed-games criterion), the
corpus realizes $\axGapJoint\pp$ against the hanab.live corpus's $\axOursGapJointGameCl\pp$ under the
same game-level clustering: above every level of the OBL hierarchy, and
$\axMinusOursGameCl\pp$ from the hanab.live corpus. The category structure replicates more closely than the aggregate: the frozen-category
profile correlates with the hanab.live corpus at $r = \axPearsonVsOursExclZero$ (excluding the
construction-zero cells), the frozen tree transfers at
$R^2 = \axRTwoStrictOblFiveTree$ (hanab.live $\hhTransferRsqStrictTree$), and indirect reference has gaps above every agent value ($\axChanIndirectColor$ for color, $\axChanIndirectRank\pp$ for rank), as do plays of unhinted newest cards ($\axChanUnhintedNewest\pp$; Table~\ref{tab:silencesplit}).

The differences are consistent with AH2AC2's selection rule (minimum score $\axScoreFloorMin$, no strikeouts; mean score $\axMeanScore$ against $\hhMeanScore$ for the hanab.live corpus). Excluding struck-out games removes games with three failed plays, which raises the aggregate gap mechanically. The rule also nearly eliminates unhinted plays (share $\axUnhintedShareTotal\%$ against $\hhUnhintedShareCombined\%$) and tilts the indirect-reference cell toward color-known cards. \emph{On the sub-population the two corpora share, the cell replicates closely} ($\axColorKnownSubcellGapPt$ against $\tTwelveAColorOnlyGap\pp$ on color-known cards). The category structure and the indirect-reference excess are therefore properties of the community's play, not of how the hanab.live corpus was constructed. Population-level independence is a different question and stays
open (\S\ref{sec:discussion}).

\subsection{The aggregate view on the OBL ladder}

The pair-clustered human corpus gap ($\hhGapJointTwoTier\pp$) lies
above every level of the OBL ladder reported in the prior paper \citep{paper1}, which runs from
$\gapOBLOnePt\pp$ at level~1 to $\gapOBLFivePt\pp$ at level~5 (Table~5
there). This is the aggregate picture behind the top row of \maintext Fig.~\ref{fig:shared}.
These are differences of aggregate gaps.
\ifdefined\appendixCarriesMovedFigs
They are not the per-cell distances of \maintext Fig.~\ref{fig:leveldist}, which average the
absolute human--level difference over the shared cells and can order the levels
differently.
\fi

\subsection{The agent-grown taxonomy transfers to people}
\label{app:transfer}

\begin{figure*}[tp]
  \centering
  \includegraphics[width=\textwidth,height=0.8\textheight,keepaspectratio]{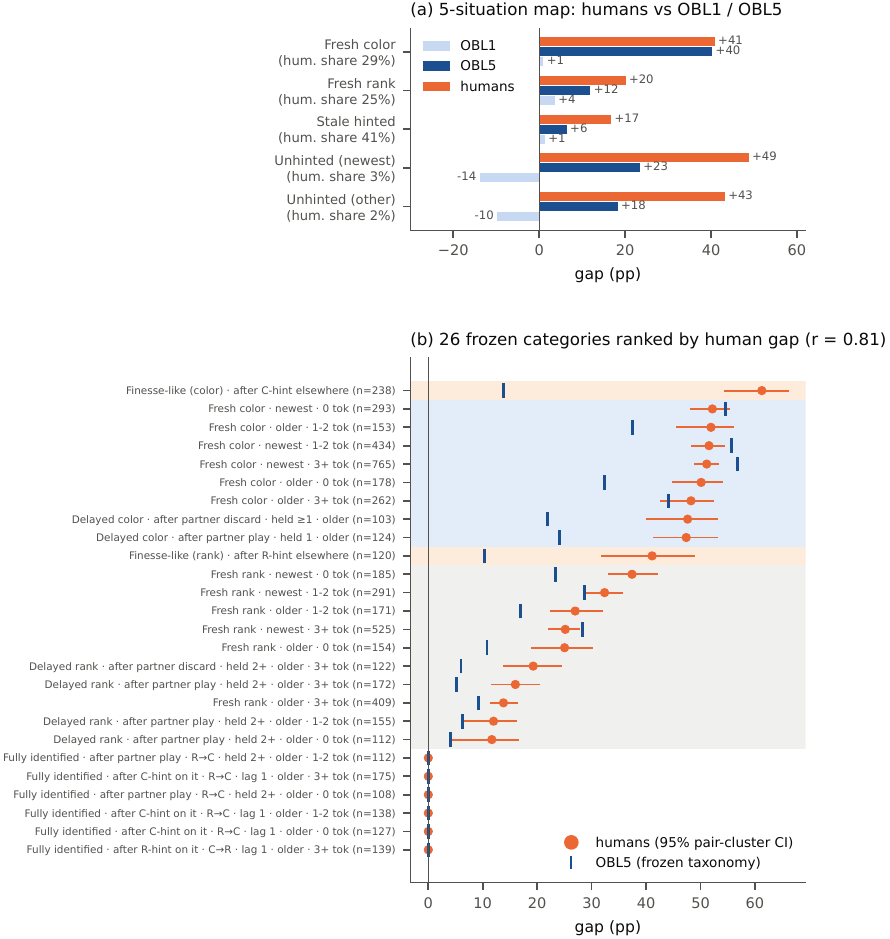}
  \caption{\textbf{Human and agent gaps on the hand-crafted situations and the frozen
  level-5 categories.} (a)~The five hand-crafted situations for humans against
  levels~1 and~5 (point estimates; the values of \maintext Table~\ref{tab:usage}, where
  ``Stale hinted'' is called ``Hinted earlier''). ``Fresh color'' and ``Fresh rank'':
  played right after a color or rank hint that touched the card; ``Stale hinted'': hinted
  earlier; ``Unhinted (newest)'' and ``Unhinted (other)'': no hint touched the card,
  newest or older card; ``hum.\ share'': share of human plays. (b)~The frozen level-5
  categories ranked by human gap: dots are humans with 95\% pair-clustered intervals,
  ticks are level~5. The $r$ in the panel title includes the $\hhTransferZeroCells$
  categories that are zero for both populations by construction (\S\ref{app:transfer}).
  Row labels follow the key of Fig.~\ref{fig:taxonomy}; $n$ is the number of human plays.
  Background shading groups rows by information family: indirect reference (orange),
  color known (blue), rank known (gray); fully identified rows are unshaded.}
  \label{fig:channels}
\end{figure*}

We freeze \S\ref{sec:where}'s level-5 machinery (discovery-half category
definitions, category means and the fitted tree, verbatim) and apply it to human
plays. The frozen
tree explains $R^2 = \hhTransferRsqStrictTree$ of human per-play residual variance out of
sample ($\hhTransferRsqPartitionTree$ using the frozen partition with human cell means),
against $\hhTransferRsqHumanTree$ for a tree grown on human data itself and
$\hhTransferRsqFiveSituations$ for the hand-crafted map. Across the $\hhTransferNumCats$
frozen categories with at least $100$ human plays, human and agent gaps correlate at
$r = \hhTransferPearsonExclZeroCells$ once the $\hhTransferZeroCells$ categories that
are zero for both sides by construction are excluded ($r = \hhTransferPearson$ with
them), with human points on or above the diagonal (Fig.~\ref{fig:channels}b).

This transfer is consistent with the metric tracking play rather than how agent games are
generated. It is not independent validation: a category's gap cannot exceed its mean
literal failure probability, and the same hint context can carry similar literal risk in
both populations, so part of the correlation may reflect these bounds.

\paragraph{Pair experience and zero-shot coordination.} The gap moves little with how often a pair has played together within the corpus (\S\ref{app:community}): near-strangers reach $\hhExpNearStrangersGap\pp$,
against $\gapOBLOneCross\pp$ for level-1 cross-play, pooled over all pairings among five
independently trained level-1 models \citep{paper1}.
Both are zero-shot coordination, of opposite kinds. Level~1 is trained to read past
actions only for their literal content, and its gap is similar in self-play and in
cross-play. Human players appear to arrive already sharing a taught system. In point estimate, the self-play gaps
of the OBL levels lie between the level-1 cross-play and near-stranger values, and
level~5's gap ($\gapOBLFivePt\pp$) lies inside the near-stranger interval.

%% file: appendix/E_depth.tex
\section{Depth of reasoning}
\label{app:depth}

On the $\chSharedNCells$ categories where the human value lies within the levels' range,
the agent profile saturates by level~2--3 and humans land at that plateau. Where humans
diverge they exceed \emph{every} level, the opposite pattern from limited recursion. A
pre-registered cognitive-hierarchy fit leaves depth unidentified ($\tau$ ran to the grid
edge in both models) and instead recovers a flat uplift of $\chBetaMTwo\pp$ on the
categories where humans exceed every level (\S\ref{app:depth:fit}).

\subsection{Distance to the human profile per level}
\label{app:leveldist}

On the shared taxonomy of \S\ref{sec:where} (estimation halves, the basis of
\maintext Fig.~\ref{fig:shared}), we compute for each level $k$ the mean per-cell absolute
difference to the human profile, $D_k = \operatorname{mean}_c\,|h(c) - g_k(c)|$,
with 95\% intervals from a joint bootstrap ($10{,}000$ draws; humans resampled by
pair, each agent corpus by game, all six corpora independently per draw; contrasts
formed within draws; recorded seed). Over all $\ldNSharedCells$ shared cells,
$D_1$ through $D_5$ read $\ldDAllOne$, $\ldDAllTwo$, $\ldDAllThree$,
$\ldDAllFour$ and $\ldDAllFive\pp$. Excluding the $\ldNExcludedChannelCells$
cells where the human value exceeds every level's with bootstrap probability at
least $0.87$ (the rule of \S\ref{app:depth:nearest}, applied to the shared cells), the
remaining core cells read $\ldDCoreOne$, $\ldDCoreTwo$,
$\ldDCoreThree$, $\ldDCoreFour$ and $\ldDCoreFive\pp$. The contrasts localize
the approach to the human profile in the first step: on the core cells
$D_1 - D_2 = \ldDCoreOneMinusTwo\pp$ against $D_2 - D_5 = \ldDCoreTwoMinusFive\pp$,
an interval that includes zero; over all cells $D_2 - D_5 = \ldDAllTwoMinusFive\pp$,
so the residual decline past level~2 is carried by the excluded cells. A signed,
human-share-weighted sensitivity gives the same shape.

\subsection{The nearest-level bootstrap}
\label{app:depth:nearest}

For each of the $\chNonzeroCats$ non-zero frozen categories (human $n \geq 100$;
$\hhTransferZeroCells$ construction-zero cells dropped), the level profile
$g_1(c), \ldots, g_5(c)$ uses each level's taxonomy corpus (one seed model at level~1,
five pooled at levels 2--5) and the human value comes from the
full corpus. The nearest-level statistic
$\arg\min_k |g_k(c) - h(c)|$, recorded as ``${>}5$'' when the human value exceeds
every level's, is bootstrapped with $\chBootDraws$ draws, resampling games in all six
corpora independently per draw. Human resampling here is \emph{game}-clustered by
design, for comparability with the agent corpora (a pair-clustered sensitivity is in
\S\ref{app:depth:paircl}). $\chTaughtCats$ of the $\chNonzeroCats$ categories exceed every
level with bootstrap probability at least $0.87$; we call them the
\emph{above-every-level} categories, and they define the indicator of the mixture fit.

\subsection{The pre-registered mixture fit}
\label{app:depth:fit}

This fit asks a simple question: can the human category profile be written as a
mixture of the five level profiles, with the mixing weights set by a single depth
parameter $\tau$ (how many steps of reasoning a typical human uses)? If depth were
what separates humans from the OBL hierarchy, the fit would return a well-identified
$\tau$ and no residual. The outcome is that $\tau$ is not identified (it runs to the
edge of its grid) and that a flat uplift on the above-every-level categories fits the
data instead.

The specification was fixed before the fit:
\begin{itemize}
  \item \textbf{M1} regresses the human category gaps on a truncated-Poisson$(\tau)$
        mixture of the five level profiles (pmf renormalized on $k \in \{1,\dots,5\}$).
  \item \textbf{M2} adds a flat uplift $\beta$ on the above-every-level categories.
  \item \textbf{Weights} are inverse bootstrap variances of the human gaps.
  \item \textbf{Fitting}: $\tau$ by grid search over $[0.05, 8]$ at step $0.01$, with
        $\beta$ in closed form given $\tau$.
  \item \textbf{Uncertainty}: a joint bootstrap ($\chBootDraws$ draws, recorded seeds)
        resampling all six corpora and refitting both models per draw.
  \item \textbf{Pre-registered truncation flags}: $\hat\tau$ at the grid edge, and
        Poisson weight on level~5 above $0.4$ at $\hat\tau$.
\end{itemize}

Both flags were raised. $\hat\tau$ ran to the grid edge in $\chEdgeDrawsMOne$ draws under M1
and $\chEdgeDrawsMTwo$ under M2 (weight on level~5 at the edge: $\chWFiveAtTau$), so
no depth estimate is quoted from the global fit; the reporting rule fixed in advance
was to state this rather than to quote an edge value. What the fit identifies is the
uplift: $\beta = \chBetaMTwo\pp$, with
$\text{wR}^2 = \chWrTwoMTwo$ (M2) against $\chWrTwoMOne$ (M1) and
$\Delta\text{AICc} = \chDeltaAICc$ in M2's favor; the uplift explains
\chCultureShare{} of weighted variance against \chCHShare{} for the best
cognitive-hierarchy component.

\subsection{Pair-clustered sensitivity}
\label{app:depth:paircl}

Because the human corpus has pair structure the agent corpora lack, a sensitivity
analysis outside the pre-registered specification clusters the human side by pair
($\hhPairsTwoTier$ pairs; recorded seeds) in both the nearest-level bootstrap and the fit
(weights and joint bootstrap). Every category keeps its most likely nearest-level
answer, and $\chTaughtCatsPairCl$ rather than $\chTaughtCats$ categories pass the $0.87$
threshold; the fit keeps the original set. Every qualitative conclusion of the fit is
unchanged: $\hat\tau$ remains at the grid edge, the uplift term remains warranted, and the
uplift moves to $\chBetaPairCl\pp$ and explains $\chCultureSharePairCl\%$ of weighted
variance (M1/M2 weighted $R^2$ $\chWrTwoMOnePairCl$/$\chWrTwoMTwoPairCl$). The $\beta$
point value and the variance split therefore depend on the weighting, and are always
quoted with this sensitivity alongside.

\subsection{The post-hoc shared-cells fit}

Restricting the pure mixture to the $\chSharedNCells$ categories where the two
populations share a convention (those not above every level) yields an interior minimum at
$\tau = \chSharedTau$ (weighted $R^2 = \chSharedWrTwo$; probability of the grid edge
$\chSharedPedge$; probability that $\tau$ falls in the literature's $1$--$3$ range
$\chSharedPinOneThree$; weight on level~1 about \chSharedWOne{}). This analysis is
\emph{post-hoc}: it was not in the pre-registered specification, rests on
$\chSharedNCells$ categories, and no claim in the paper depends on it. It is reported because
it makes clear what can be identified: depth can be estimated on the
shared conventions and cannot be estimated off them.

\subsection{What is and is not identified}

Two questions must be kept apart. \emph{Which single level is a category nearest to?}
is answered by the nearest-level bootstrap (\S\ref{app:depth:nearest}), category by
category, and for $\chTaughtCats$ of them the answer is ``above all of them.'' \emph{What mixture of levels
best reproduces the human profile?} is the parametric question, and its depth parameter
is unidentified globally because these categories exceed the entire basis: no
mixture of levels reaches them, so the fit pushes $\tau$ to the grid edge and the flags
fire. The identified content of the exercise is the decomposition, a shared component
saturating by level~2--3 plus an uplift no level supplies, together with the
caveats that the basis truncates at level~5 and that each level is trained with beliefs
induced by a fixed base policy (uniformly random at level~1, the previous level's policy
above it). The levels are therefore not literal cognitive-hierarchy levels.

\subsection{Depth is not what separates them: in full}
\label{sec:depth}

\begin{figure*}[tp]
  \centering
  \includegraphics[width=\textwidth,height=0.8\textheight,keepaspectratio]{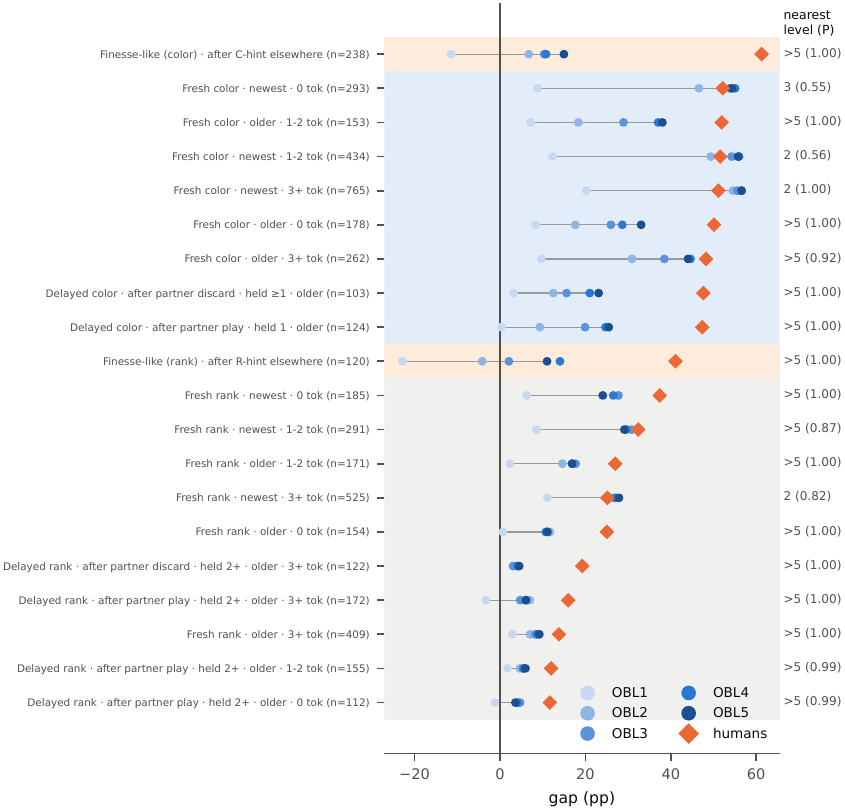}
  \caption{\textbf{Per-category convention gap across levels 1--5 with the human value
  overlaid.} Rows are the $\chNonzeroCats$ non-zero frozen level-5 categories with at
  least $100$ human plays; the row labels follow the key of Fig.~\ref{fig:taxonomy} and $n$ is the number of human plays,
  and the shading groups rows by information family as there. Markers give each level's
  gap and the human value. The right column gives the level nearest to the human value
  and its bootstrap probability; ``${>}5$'' means the human value exceeds every level.
  Where humans and agents share a convention, the agent profile plateaus by level~2--3
  and humans land at that plateau; where humans diverge they exceed every level.}
  \label{fig:depth}
\end{figure*}

Behavioral game theory finds that people use one to three steps of recursive reasoning
about others \citep{stahl1995,nagel1995,camerer2004ch}, which appears to conflict with a human gap above every level of a five-level
hierarchy. Looking category by category removes the conflict (Fig.~\ref{fig:depth}).

On the three newest-card fresh-color categories (three of the $\chSharedNCells$ whose
nearest level is 2 or 3), the agent profile \emph{saturates}: level~2 already reaches
\chOBLTwoCoreRange\pp{} and levels 3--5 add almost nothing. Humans land at
\chHumanCoreRange\pp: at the level-2--3 value, and below the level-5
asymptote ($P(\text{human} < \text{level 5}) =$ \chPbelowFiveCoreRange{}).

Where humans diverge they are \emph{above every level} ($\chTaughtCats$ of
$\chNonzeroCats$ non-zero categories), which is the opposite pattern from limited
recursion. Those categories split into two shapes: profiles still rising at level~5, and flat-low
profiles, including both indirect-reference categories, that stay low at every level.

Neither side needs to recurse at the table. The agents' recursion is performed during training and folded into a deployed
policy that acts without search; taught human conventions may be a community's reasoning
compiled into lookup rules over years. The gap describes the \emph{product}, how far play departs from the literal prediction, not
the \emph{process}. High human values therefore do not contradict $k$-level limits,
which concern on-the-fly reasoning in novel situations.

%% file: appendix/F_repro.tex
\section{Reproducibility}
\label{app:repro}

This appendix states how every number this paper computes can be regenerated.

\paragraph{Metric engine.}
All posteriors are computed by the metric implementation released with the previous study \citep{paper1},
used at a pinned commit rather than as an installed package, so the joint posterior
here is bit-identical to the previous study's (verified play-for-play on the human corpus; Appendix~\ref{app:corpus}).

\paragraph{Replay verification.}
Every game entering any analysis, generated or observational, is replayed in the
metric engine and required to reproduce its ground-truth final score exactly before any
statistic is computed. This holds for the self-play and mixed-level conditions
(\S\ref{sec:setup}), the seed sweep ($\tSevenGamesTotal$ games), every intervention
condition and its baseline (Appendix~\ref{app:interventions}), the human corpus, the
AH2AC2 release ($\axNGamesReleased/\axNGamesReleased$), and every clone evaluation
(\S\ref{sec:proxies}). No condition reported in this paper has a replay mismatch.

\paragraph{The audit pipeline.}
The OBL ladder values (\S\ref{sec:hierarchy})
are quoted as reported in the previous study \citep{paper1}; this paper's recomputation
from the same games gives the same point values, and some interval ends differ by
hundredths of a percentage point. Each analysis emits audited entries (value, interval,
source artifact path, recomputation script, clustering and seed) into versioned JSON
files; a generator turns those entries into the numbers the manuscript prints, and the
build fails if any printed number lacks an audited entry.
Every entry records the pipeline artifact it was recomputed from, so any value this
paper computes can be traced to per-play records without consulting the manuscript itself.

\paragraph{Seed policy.}
Environment seeds define disjoint blocks per condition; paired designs share their
block with their baseline by construction. Analysis seeds (splits, bootstraps, fits)
are recorded in the audit entries alongside draw counts. One documented exception to
bit-level reproduction: the per-category bootstrap of \S\ref{sec:where} keys its seed
on the Python hash of the category label, so those intervals reproduce only up to bootstrap error
rather than bit-for-bit; every other interval this paper computes reproduces exactly from
its recorded seed.

\paragraph{Release.}
\releasesentence{} The human and AH2AC2 corpora are redistributed only as derived,
de-identified records consistent with their sources' terms. Raw agent game logs stay on the generation server, because they derive from checkpoints licensed for non-commercial use; the included generation scripts regenerate them. The generation records name each checkpoint and environment seed but no code commit or checkpoint hash. The OBL commit and the hashes of all $25$ checkpoints used here were read from the generation server in October 2026 and describe the files at that time. On that server, the unmasked intervention code path regenerated the level-5
baseline action for action ($\eFiveGateLogOnlyIdentical$ games;
Appendix~\ref{app:interventions}).\releaseanon

%% file: appendix/G_clones.tex
\section{Behavioral clones: does cloning preserve the human-specific cells?}
\label{app:clones}\label{sec:proxies}

\paragraph{Behavioral clones.}
We trained clones of human play and of level-5 play identically, at matched data
scale, with registered predictions. They reproduce the asymmetry of \S\ref{sec:what}.
The control clone preserves its source's structure
($r = \bcOCTwentyKProfileRVsOblSelf$ at scale). The human clone reaches level~5's
gap on the dominant shared convention ($\bcHCFreshColorNewestThreepXObl\pp$), while indirect reference and the unhinted newest-card cell collapse to $\bcHCChanIndirectColorXObl$ and $\bcHCChanUnhintedNewestXObl\pp$ (the latter below literal chance). Aggregate similarity to humans is not fidelity on
the channels where humans differ from machines.

The channels of \S\ref{sec:what} matter beyond this paper because the field
increasingly evaluates coordination agents against \emph{behaviorally-cloned} human
proxies rather than people; AH2AC2's own evaluation protocol rests on behavioral-cloning
stand-ins for the community measured above \citep{ah2ac2}. Cloning imitates the
distribution of human actions; whether it also reproduces the human gaps on indirect reference and on unhinted cards is what our instrument can test.

\paragraph{Design.}
We train a clone of human play (HC: the AH2AC2 open games, $1{,}700$ training /
$158$ held out; the hanab.live corpus, the human reference below, is not used for
training) and, as the control, a clone of level-5
self-play (OC). Both use identical architecture, hyperparameters, data size and
training steps: the OBL codebase's encoder and policy network, trained by supervised
learning only, and one frozen configuration for every clone. If OC preserves its source's
convention structure while HC loses the human-specific cells, the loss is cell-specific
rather than an artifact of data scale. An OC series at $\axNGamesReleased$, $5{,}000$
and $20{,}000$ games calibrates what cloning captures at each scale. Four predictions were registered, with timestamps, before any training:
the level-5 clone keeps its source's profile and fresh-color gap; the human clone keeps
the fresh-color conventions but not the indirect-reference and unhinted cells;
preservation improves with data; and the human clone is less coherent in self-play than
the level-5 clone. Clones are measured with the full ground-truth pipeline of
\S\ref{sec:setup}: exact posteriors, replay verification, the frozen taxonomy. Because
matched-scale clones are not coherent in self-play (HC: $\bcHCZeroScoreRateSelf\%$
zero-score games; OC: $\bcOCMatchedZeroScoreRateSelf\%$), their channel numbers
come from a frozen-partner variant (clone fixed in seat~0 against level~5,
$500$ games, clone-side plays only, still fully ground-truthed).

\paragraph{Cloning works where the data are dense.}
At $20{,}000$ games the level-5 clone reproduces its source: category-profile
$r = \bcOCTwentyKProfileRVsOblSelf$ against level~5, gap $\bcOCTwentyKGapSelf\pp$
(source: $\gapOBLFive$), fresh-color cells within a point of the original
($\bcOCTwentyKFreshColorNewestThreepSelf$ vs $\tOneZeroGapThree\pp$). Preservation is
monotone in data ($r = \bcOCScalingR$ across the series). The method, the
architecture and the instrument are therefore all sufficient; what cloning captures
is a question of what the data contain and how often.

\paragraph{The shared conventions transfer; the human-specific cells do not.}
At matched scale, the human clone reaches level~5's gap on the dominant shared
convention: in the level-5 category for an immediate play after a fresh color hint on
the newest card, it reads $\bcHCFreshColorNewestThreepXObl\pp$, against level~5's
$\tOneZeroGapThree\pp$ for the same category. The cells that separate humans from the OBL hierarchy do not survive cloning: indirect reference reads
$\bcHCChanIndirectColorXObl$ (color; humans $\hhCatIndirectColorHuman$) and
$\bcHCChanIndirectRankXObl\pp$ (rank; humans $\hhCatIndirectRankHuman$), and the unhinted cells read $\bcHCChanUnhintedNewestXObl$ and $\bcHCChanUnhintedOtherXObl\pp$ against the
humans' $\hhSitUnhintedNewestGap$ and $\hhSitUnhintedOlderGap$. Every interval excludes
the human value by tens of points. The control rules out the generic explanation: at the
same data and configuration, OC preserves its source's structure
($r = \bcOCMatchedProfileRVsOblXObl$ with the frozen partner). What HC loses tracks how
\emph{often} a cell appears in training (each unhinted cell holds $\axSitUnhintedOtherShare$--$\axSitUnhintedNewestShare\%$
of AH2AC2 plays), not how large its gap is.

\paragraph{Unhinted plays invert.} On the frozen-partner runs, the clone's unhinted newest-card plays are \emph{worse than literal chance} ($\bcHCChanUnhintedNewestXObl\pp$; failure rate
$\bcHCChanUnhintedNewestLossXObl\%$), as are level~1's in self-play
($\tFiveGapUnhintedNewestOBLOne\pp$). The clone's value rests on few plays. A negative
value here is consistent with a partner whose hints favor playable cards: the cards it
leaves untouched are then less likely to be playable than the literal observer assumes.
The partner alone does not account for the human clone's value: with the same partner,
the matched-scale level-5 clone reads $\bcOCMatchedChanUnhintedNewestXObl\pp$ on this
cell. Cloning reproduces the \emph{act} of playing an unhinted card, not the situations
in which humans make that play; one reading is that imitating surface behavior without
the representation behind it is costly rather than neutral.

\paragraph{Scope.}
This measures plain behavioral cloning at open-split scale ($1{,}700$--$18{,}300$
games, default configuration, no reinforcement learning). The deployed AH2AC2 proxies, cloned from $147{,}621$ closed games and refined with
regularized self-play, are not measured here. The scaling series warns against reading
absence as impossibility: structure preservation improves with data, so channel
survival at $10^5$ games is not ruled out. These results suggest that the
shared, frequent conventions and the rare channels respond to imitation differently, so a
proxy that matches human play on the former can differ from humans on the channels where
humans differ from machines. Aggregate similarity is not channel
fidelity: the same dissociation, at the level of evaluation methodology, that
\S\ref{sec:worth} found at the level of the metric.